\documentclass[11pt,letterpaper]{article}
\usepackage[margin=1in]{geometry}
\usepackage[T1]{fontenc}
\usepackage[utf8]{inputenc}
\usepackage{newtxtext,amsmath,amsthm,mathtools,newtxmath}
\usepackage{microtype}
\usepackage{enumitem,array}
\usepackage{tikz}
\usepackage{float}
\usetikzlibrary{arrows.meta,decorations.pathreplacing,calc}
\usepackage[ruled,vlined]{algorithm2e}
\SetAlgoSkip{}
\SetAlgoNoEnd
\SetAlCapSkip{0.2em}
\SetInd{0.35em}{0.6em}
\SetKwInput{KwIn}{Input}
\SetKwInput{KwOut}{Output}
\usepackage[numbers,sort&compress]{natbib}
\usepackage[hidelinks]{hyperref}
\hypersetup{pdftitle={Reducing Asymmetric Path TSP to Asymmetric TSP},pdfauthor={Hong Li}}
\newtheorem{theorem}{Theorem}[section]
\newtheorem{lemma}[theorem]{Lemma}
\newtheorem{corollary}[theorem]{Corollary}

\theoremstyle{definition}

\theoremstyle{remark}

\newcommand{\OPT}{\mathrm{OPT}}

\newcommand{\meas}{\operatorname{meas}}
\newcommand{\up}{\operatorname{up}}
\newcommand{\Ps}{P^*}

\newcommand{\eps}{\varepsilon}
\newcommand{\cU}{\mathcal{U}}
\newcommand{\cH}{\mathcal{H}}
\newcommand{\cR}{\mathcal{R}}
\newcommand{\cP}{\mathcal{P}}
\newcommand{\Pt}{\widetilde{\mathcal P}}
\newcommand{\cPs}{\mathcal{P}^*}
\newcommand{\cQ}{\mathcal{Q}}
\newcommand{\ee}{\mathrm{e}}

\title{Reducing Asymmetric Path TSP to Asymmetric TSP}
\author{%
Hong Li\\
\small School of Mathematics and Statistics, Yunnan University\\
\small \texttt{honglimath@126.com}
}
\date{}

\begin{document}
\maketitle

\begin{abstract}
In the Asymmetric Traveling Salesman Problem (Asymmetric TSP), we are given distances between $n$ vertices that satisfy the triangle inequality but need not be symmetric, and we seek a shortest tour through all vertices. In its path version, Asymmetric Path TSP, two vertices $s$ and $t$ are prescribed, and we seek a shortest path from $s$ to $t$ through all vertices. Both problems are equivalent to their versions with arbitrary nonnegative edge costs on a directed graph, in which vertices may be visited more than once. We show that for all constants $\alpha>1$ and $\eps>0$, any $\alpha$-approximation algorithm for Asymmetric TSP yields an $(\alpha+\eps)$-approximation algorithm for Asymmetric Path TSP. Consequently, every constant-factor approximation guarantee for Asymmetric TSP carries over to Asymmetric Path TSP up to an arbitrarily small additive error. The best previous reduction of this kind, due to Feige and Singh, yields an $\alpha(2+\eps)$-approximation algorithm. Our result answers an open question of Traub, Vygen and Zenklusen. The same framework also gives such reductions for Asymmetric Multi-Path TSP with a fixed number of paths, and for unweighted digraphs. Like earlier dynamic programs for Path TSP, ours uses guessed edges of an optimum solution to improve the cost bound; unlike them, it also uses their cost to bound how many edges it has to guess.
\end{abstract}

\section{Introduction}\label{sec:intro}

\subsection{The problem and our results}\label{sec:intro-results}

In the \emph{Asymmetric Traveling Salesman Problem} (Asymmetric TSP), we are given nonnegative distances on $n$ vertices that satisfy the triangle inequality but need not be symmetric, a \emph{directed metric}, and seek a shortest tour through all vertices. In \emph{Asymmetric Path TSP}, two distinct vertices $s$ and $t$ are prescribed, and we seek a shortest path from $s$ to $t$ through all vertices. Our main result answers a question raised by Traub, Vygen and Zenklusen~\cite[Section~8]{TVZ22} and stated as Open Problem~16.30 in~\cite{TV25}.\footnote{\label{fn:lean}Theorems~\ref{thm:main}--\ref{thm:graph-path} are formally verified in Lean~4: \url{https://github.com/HongLiMath/path-atsp-lean}.}

\begin{theorem}\label{thm:main}
Let $\alpha>1$ and $\eps>0$ be constants. If there is an $\alpha$-approximation algorithm for Asymmetric TSP, then there is an $(\alpha+\eps)$-approximation algorithm for Asymmetric Path TSP.
\end{theorem}

The reduction does not depend on the internal structure of the given algorithm. Thus every constant-factor approximation guarantee for Asymmetric TSP carries over to its path version up to an arbitrarily small additive error. For example, with the $(R^*+\eps)$-approximation algorithm of Hajiaghayi and Yun~\cite{HY26} for Asymmetric TSP, where $R^*=6+\sqrt2+\sqrt{15+6\sqrt2}\approx12.26$, it gives an $(R^*+\eps)$-approximation algorithm for Asymmetric Path TSP, matching the guarantee that they obtained directly for the path version; any future improvement for Asymmetric TSP carries over in the same way. The best previous general reduction, due to Feige and Singh~\cite{FS07}, gives $\alpha(2+\eps)$. Our reduction makes polynomially many calls to the given algorithm and performs polynomial additional work; the degrees of these polynomials depend only on $\alpha$ and $\eps$. The same algorithm also applies to symmetric distances, where the two problems become metric TSP and Path TSP. There it gives another reduction with the guarantee $\alpha+\eps$ (Corollary~\ref{cor:symmetric}); such reductions were given by Traub, Vygen and Zenklusen~\cite{TVZ22} and by Traub and Vygen~\cite[Corollary~16.25]{TV25}.

\paragraph{Main idea.}
Our algorithm is a dynamic program that guesses the edges by which an optimum solution connects the parts of a partition of the vertex set; inside each part, it uses a tour computed by the given algorithm for Asymmetric TSP. Guessed edges are paid for at their own cost, whereas the parts are covered only approximately. A finer partition can guess more optimum edges and may tolerate larger errors inside the parts; if every part is a single vertex, all edges of the optimum solution are guessed and the solution is recovered exactly. But a finer partition may also require guessing more crossing edges per set, which makes the dynamic program larger.

Guessing the edges of an optimum solution that cross certain vertex sets is also at the core of the reductions for Path TSP~\cite{TVZ22,TV25}. Starting from a constant-factor approximation algorithm, these reductions reach $\alpha+\eps$ after a constant number of rounds. To keep each dynamic program polynomial, they restrict the vertex sets considered and how often the optimum solution may cross them, with bounds fixed in advance for each round. Their analysis uses the symmetry of the distances to guarantee, where needed, that the guessed edges make up a large part of the optimum cost; since these edges are paid for at their own cost, each round closes part of the remaining gap to $\alpha+\eps$. In a directed metric, the guessed edges need not make up a large part of the optimum cost, so the same argument no longer guarantees an improvement in each round.

We control the size of the dynamic program differently. Our analysis refines the partition step by step until the error inside the parts can be tolerated. This error is bounded in terms of the largest distances inside the parts in either direction, so controlling it also handles the asymmetry of the distances.

Our analysis divides vertex sets repeatedly, each time by the distances from one vertex of the set. A division can increase the number of crossing edges to be guessed per set, and we control this growth by cost: the growth at each step is paid for by the cost of the edges that this step turns into edges between parts, relative to the error inside the parts that can currently be tolerated. Since an edge becomes an edge between parts at most once, a global accounting bounds the number of crossing edges that need to be guessed per set by a constant that depends only on the fixed parameters. Unlike the per-round bounds of the reductions above, this bound holds throughout the refinement. Our dynamic program considers a family of vertex sets that contains all sets arising in this analysis, and it makes these guesses in polynomial time. A single run reaches $\alpha+\eps$. Section~\ref{sec:overview} explains this approach in more detail.

\paragraph{More general results.}
In \emph{Asymmetric Multi-Path TSP}, $k$ ordered pairs $(s_1,t_1),\ldots,(s_k,t_k)$ are given, and we seek paths from $s_i$ to $t_i$, one for each $i$, that together visit every vertex, at minimum total cost; the paths may share vertices, and for $s_i=t_i$ the $i$th path is a tour containing $s_i$.\footnote{Our results for this problem also apply when a fixed number of prescribed vertices must occur in a specified cyclic order on a tour through all vertices of a directed metric, a problem studied in a more general form by Chekuri and P\'al~\cite{CP07}: the order is encoded by the pairs of consecutive prescribed vertices, and concatenating the paths of a solution in this order and shortcutting gives a tour with the required order.}

\begin{theorem}\label{thm:unified}
Let $\alpha>1$, $\eps>0$, and $k\ge1$ be constants. An $\alpha$-approximation algorithm for Asymmetric TSP yields an $(\alpha+\eps)$-approximation algorithm for Asymmetric Multi-Path TSP with $k$ prescribed pairs.
\end{theorem}

Our reduction also applies to unweighted digraphs. \emph{Asymmetric Graph TSP} and \emph{Asymmetric Graph Path TSP} are the special cases of Asymmetric TSP and Asymmetric Path TSP in which the metric is the \emph{graph metric} of a strongly connected unweighted digraph, that is, its shortest-path distances~\cite{TV25}.

\begin{theorem}\label{thm:graph-path}
Let $\alpha>1$ and $\eps>0$ be constants. An $\alpha$-approximation algorithm for Asymmetric Graph TSP yields an $(\alpha+\eps)$-approximation algorithm for Asymmetric Graph Path TSP.
\end{theorem}

Together with the algorithms of Vygen~\cite{Vyg26} for Asymmetric Graph TSP, this gives a $(3+2\sqrt2+\eps)$-approximation algorithm for Asymmetric Graph Path TSP for every $\eps>0$, where $3+2\sqrt2\approx5.83$ (Section~\ref{sec:applications}). In concurrent work, Christalla and Traub~\cite{CT26} obtained an LP-relative $(5+2\sqrt2+\eps)$-approximation algorithm for this problem. All our results follow from one algorithm, whose analysis Section~\ref{sec:graphic} adapts to unweighted digraphs. Besides the given algorithm for Asymmetric TSP, it uses a rough solution: a bounded number of walks that together visit every vertex and cost at most a constant times the optimum. The dynamic program uses this solution to decide which vertex sets to consider. For Asymmetric Path TSP, one run of the Feige--Singh reduction with error parameter~$1$ provides such a solution, namely an $s$-$t$ path of cost at most $3\alpha$ times the optimum. A reader interested only in Theorem~\ref{thm:main} may keep in mind one prescribed pair $(s,t)$ and this single path throughout. Appendix~\ref{sec:initialization} constructs such solutions for the other results.

\subsection{Related work}\label{sec:related}

\paragraph{Asymmetric TSP.}
Feige and Singh~\cite{FS07} obtained an approximation ratio of $\frac23\log_2n$, and Asadpour, Goemans, M\k{a}dry, Oveis Gharan and Saberi~\cite{AGMOS17} a ratio of $O(\log n/\log\log n)$. Svensson~\cite{Sve15} found the first constant-factor approximation algorithm for Asymmetric Graph TSP, and Svensson, Tarnawski and V\'egh~\cite{STV20} the first one for general instances. Traub and Vygen improved the ratio to $22+\eps$~\cite{TV22} and then to $17+\eps$~\cite{TV25}. Vygen obtained $12+2\sqrt2+\eps$, and hence a ratio below $15$; a further refinement gives a ratio below $14.7$~\cite{Vyg26}. Recently, Hajiaghayi and Yun~\cite{HY26} obtained $R^*+\eps$, where $R^*=6+\sqrt2+\sqrt{15+6\sqrt2}\approx12.26$.

\paragraph{Asymmetric Path TSP.}
Lam and Newman~\cite{LN08} obtained an approximation ratio of $O(\sqrt n)$, and Chekuri and P\'al~\cite{CP07} one of $O(\log n)$. Feige and Singh~\cite{FS07} showed that any $\alpha$-approximation algorithm for Asymmetric TSP yields an $\alpha(2+\eps)$-approximation algorithm for Asymmetric Path TSP. Through this reduction, every improvement for Asymmetric TSP carried over to the path version, but the resulting ratio was more than twice as large. Subsequent algorithms for Asymmetric Path TSP obtained ratios of $17+\eps$ by Traub and Vygen~\cite{TV25}, $15$ by Vygen~\cite{Vyg26}, and $R^*+\eps$ by Hajiaghayi and Yun~\cite{HY26}. These guarantees are relative to the natural LP relaxation and therefore also bound its integrality ratio. An $\alpha$-approximation algorithm for Asymmetric Graph TSP yields a $(2\alpha-1)$-approximation algorithm for Asymmetric Graph Path TSP~\cite[Theorem~9.11]{TV25}, building on work of K\"ohne, Traub and Vygen~\cite{KTV20}.

\paragraph{Several paths.}
Friggstad~\cite{Fri13} studied Asymmetric Multi-Path TSP. He obtained an $O(\log n)$-approximation for two pairs and showed that, when the number of pairs is part of the input, no finite approximation ratio is possible unless $\mathrm{P}=\mathrm{NP}$. In Theorem~\ref{thm:unified}, the number of pairs is fixed. Christalla and Traub~\cite{CT26} obtained an $O(k)$-approximation algorithm for Asymmetric Multi-Path TSP with running time $n^{O(k)}$.\footnote{Theorem~\ref{thm:unified} answers their Open Question~4.1, which asks whether a ratio that does not depend on $k$ can be achieved in time $n^{g(k)}$ for some function $g$.}

\paragraph{The symmetric case.}
When the distances are symmetric, the tour and path problems become metric TSP and Path TSP. For Path TSP, Traub~\cite{Tra17}, inspired by the work of Blum et al.~\cite{BCKLMM07}, introduced a dynamic program that guesses edges of an optimum solution. This technique was developed further by Traub and Vygen~\cite{TV19a} and by Zenklusen~\cite{Zen19}. Building on it, Traub, Vygen and Zenklusen~\cite{TVZ22} showed that, for constants $\alpha>1$ and $\eps>0$, any $\alpha$-approximation algorithm for metric TSP yields an $(\alpha+\eps)$-approximation algorithm for Path TSP, also for unweighted graphs. Together with Karlin, Klein and Oveis Gharan~\cite{KKO24}, this gives a ratio below $\frac32$ for Path TSP.

Traub and Vygen~\cite{TV25} gave a different proof, which also covers Multi-Path TSP, the symmetric version of Asymmetric Multi-Path TSP, with a fixed number of terminal pairs. Starting from a constant-factor approximation algorithm for Multi-Path TSP, their algorithm improves the ratio over a constant number of rounds. Each round runs two algorithms and returns the better solution. The first combines tours computed by the given algorithm for metric TSP with shortest paths between the terminal pairs. The second is a dynamic program that guesses the edges of an optimum solution in certain cuts and solves the parts between these cuts with the algorithm for Multi-Path TSP from the previous round. The analysis distinguishes two cases. If the total distance between the terminal pairs is small, the first solution is good. If it is large, the guessed edges make up a large fraction of the optimum cost; since they are paid for at their own cost instead of approximately, the second solution improves on the previous ratio. Corollary~\ref{cor:symmetric} gives another reduction with the same guarantees, using a single dynamic program. In previous work~\cite{Li26}, we gave a reduction of the same kind from the path version of Prize-Collecting TSP to Prize-Collecting TSP.

\subsection{Overview of the approach}\label{sec:overview}

We explain the reduction for Theorem~\ref{thm:main}, that is, for one prescribed pair $(s,t)$ in a directed metric; Section~\ref{sec:prelim} describes the setting with several prescribed pairs, and Section~\ref{sec:graphic} the changes for unweighted digraphs. We regard the directed metric as a complete directed graph on its vertex set $V$, in which the edge $(a,b)$ costs $c(a,b)$, the distance from $a$ to $b$; for a set $F$ of edges, $c(F)$ is its total cost. An $s$-$t$ path or walk is one from $s$ to $t$, and a tour of a vertex set $U$ is a tour whose vertices are exactly those of $U$. Fix an optimum $s$-$t$ path $\Ps$ through all vertices, and let $\OPT$ be its cost. The algorithm does not know $\Ps$, which is used only in the analysis. We call the given $\alpha$-approximation algorithm for Asymmetric TSP the \emph{oracle}.

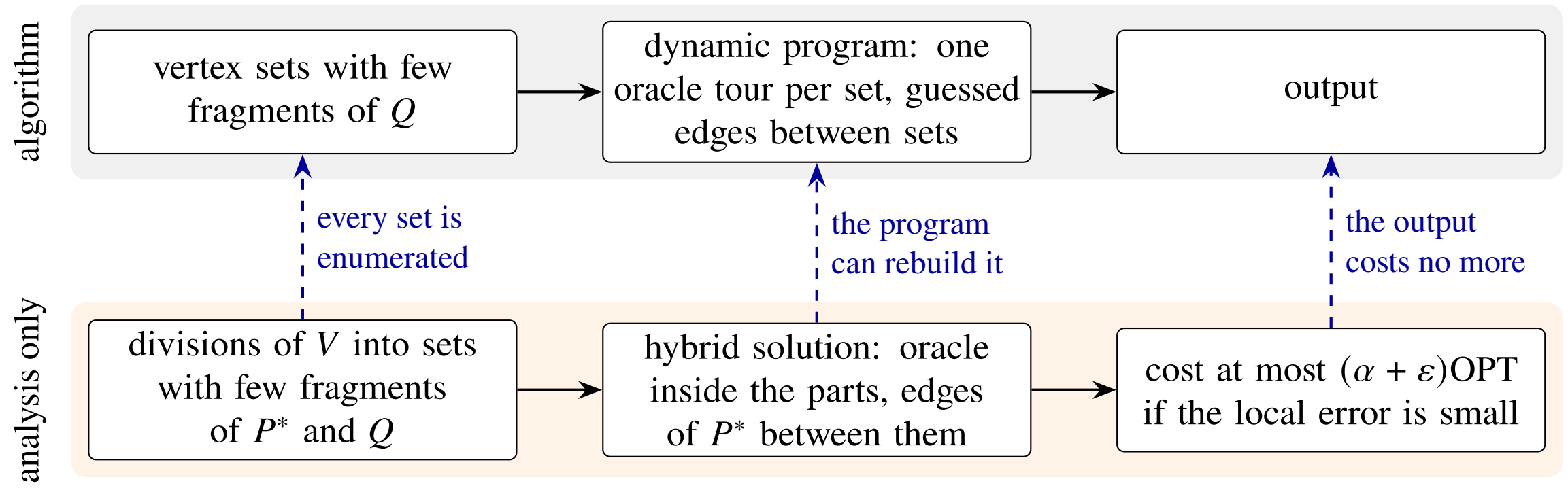
\begin{figure}[htbp]
\centering
\begin{tikzpicture}[>=Stealth,
  box/.style={draw,rounded corners=2pt,align=center,text width=3.95cm,minimum height=1.2cm,font=\small,fill=white,inner sep=3pt},
  link/.style={->,thick},
  cmp/.style={->,thick,dashed,blue!60!black}]
  \fill[gray!12,rounded corners] (-2.25,0.85) rectangle (12.25,-0.85);
  \fill[orange!10,rounded corners] (-2.25,-2.05) rectangle (12.25,-3.75);
  \node[rotate=90,font=\small] at (-2.65,0) {algorithm};
  \node[rotate=90,font=\small] at (-2.65,-2.9) {analysis only};
  \node[box] (a1) at (0,0) {vertex sets with few fragments of $Q$};
  \node[box] (a2) at (5,0) {dynamic program: one oracle tour per set, guessed edges between sets};
  \node[box] (a3) at (10,0) {output};
  \node[box] (b1) at (0,-2.9) {divisions of $V$ into sets\\ with few fragments\\ of $\Ps$ and $Q$};
  \node[box] (b2) at (5,-2.9) {hybrid solution: oracle inside the parts, edges of $\Ps$ between them};
  \node[box] (b3) at (10,-2.9) {cost at most $(\alpha+\eps)\OPT$ if the local error is small};
  \draw[link] (a1)--(a2); \draw[link] (a2)--(a3);
  \draw[link] (b1)--(b2); \draw[link] (b2)--(b3);
  \draw[cmp] (b1)-- node[right,font=\footnotesize,align=left] {every set is\\ enumerated} (a1);
  \draw[cmp] (b2)-- node[right,font=\footnotesize,align=left] {the program\\ can rebuild it} (a2);
  \draw[cmp] (b3)-- node[right,font=\footnotesize,align=left] {the output\\ costs no more} (a3);
\end{tikzpicture}
\caption{The objects of the proof. Only the top row is computed: the algorithm knows the distances, the oracle, and $Q$. The bottom row depends on the unknown optimum $\Ps$ and is used only in the analysis. The dashed arrows are the comparisons between the rows (Lemma~\ref{lem:first-order} and Theorem~\ref{thm:dp}).}
\label{fig:framework}
\end{figure}

The overview follows how the algorithm is designed. Like the dynamic programs for Path TSP, our algorithm applies an approximation algorithm, here the oracle, inside vertex sets and guesses the edges between them. We first explain why asymmetric distances call for a fine division of the vertex set, and bound the error inside a single set. This bound yields a condition under which a partition of $V$ gives a good solution, and the condition determines what the dynamic program has to guess. It then remains to show, in the analysis, that such a partition can be reached by divisions that keep the dynamic program polynomial. Figure~\ref{fig:framework} at the end of this overview shows which objects the algorithm computes and which exist only in the analysis.

\paragraph{The difficulty of asymmetric distances.}
The simplest way to use the oracle is to call it on the whole instance, traverse the returned tour from $s$, and then go from $s$ to $t$. To bound the cost of the returned tour, close $\Ps$ with the edge $(t,s)$; this gives a tour of cost $\OPT+c(t,s)$, so the returned tour costs at most $\alpha$ times this. Adding the edge $(s,t)$, the resulting $s$-$t$ walk costs at most
\[
 \alpha\OPT+\alpha c(t,s)+c(s,t),
\]
which is good when both distances between $s$ and $t$ are small.

When the distance between $s$ and $t$ is large, the proofs for the symmetric case described in Section~\ref{sec:related} rely on the guessed edges instead: a large total distance between the terminal pairs forces them to make up a large fraction of the optimum cost. In a directed metric, a large $c(t,s)$ does not force this, so the argument no longer guarantees that a round improves the bound. Traub, Vygen and Zenklusen~\cite{TVZ22} note that in this case they do not know how to guess edges of significant cost by dynamic programming. The bound above has an analogue inside a vertex set, with the largest distance in the set, which bounds the distances in both directions, in the role of $c(t,s)$ and $c(s,t)$. Our approach divides the vertex set finely, which can make these largest distances small, while keeping the dynamic program polynomial.

\paragraph{Working inside a vertex set.}
We first make this precise for a single vertex set. For a set $U$ of vertices, let $d_U=\max_{u,v\in U}c(u,v)$ be its diameter, taken over ordered pairs. The maximal subpaths of $\Ps$ inside $U$ are the \emph{fragments} of $\Ps$ in $U$; a fragment may consist of a single vertex. Write $r_{\Ps}(U)$ for their number. Each fragment starts at an entry and ends at an exit, and we keep each entry paired with its exit. In the analysis, we replace the fragments by walks inside $U$ that have the same entry--exit pairs and together visit all of $U$. The rest of $\Ps$ is unchanged, so the result is again an $s$-$t$ walk that visits every vertex.

\begin{figure}[htbp]
\centering
\begin{tikzpicture}[scale=0.94,>=Stealth,
  vtx/.style={circle,fill=black,inner sep=1.4pt},
  port/.style={circle,draw,fill=white,inner sep=1.7pt},
  arc/.style={thick,->,shorten >=2pt,shorten <=2pt},
  add/.style={blue,dashed,thick,->,shorten >=2pt,shorten <=2pt}]
  \foreach \shift/\heading in {0/{Closing the fragments},7/{The local solution}}{
    \begin{scope}[xshift=\shift cm]
      \node at (1.8,3.25) {\heading};
      \node[port,label=left:$a_1$] (s1) at (0,2.1) {};
      \node[vtx] (x) at (1.8,2.7) {};
      \node[port,label=right:$b_1$] (t1) at (3.6,2.1) {};
      \node[port,label=right:$a_2$] (s2) at (3.6,0.1) {};
      \node[vtx] (y) at (1.8,-0.5) {};
      \node[port,label=left:$b_2$] (t2) at (0,0.1) {};
      \ifnum\shift=0
        \draw[arc] (s1)--(x); \draw[arc] (x)--(t1);
        \draw[arc] (s2)--(y); \draw[arc] (y)--(t2);
        \draw[add] (t1)-- node[right] {$\le d_U$} (s2);
        \draw[add] (t2)-- node[left] {$\le d_U$} (s1);
        \node at (1.8,1.1) {$\Ps[U]$};
      \else
        \draw[arc] (s1)--(t2); \draw[arc] (t2)--(y);
        \draw[arc] (y)--(s2); \draw[arc] (s2)--(t1);
        \draw[arc] (t1)--(x); \draw[arc] (x)--(s1);
        \draw[add] (s1)-- node[below] {$\le d_U$} (t1);
        \draw[add] (s2)-- node[above] {$\le d_U$} (t2);
        \node at (1.8,1.1) {\small oracle tour};
      \fi
    \end{scope}
  }
\end{tikzpicture}
\caption{The two uses of the diameter. Left: the two fragments of $\Ps$ in $U$, from $a_1$ to $b_1$ and from $a_2$ to $b_2$, are linked into a tour of $U$, which bounds the cost of the oracle tour. Right: the local solution. The first walk traverses the oracle tour from $a_1$ and then takes the edge from $a_1$ to $b_1$; the second walk is the edge from $a_2$ to $b_2$. Both walks keep the entry--exit pairs of the fragments. Dashed arrows are added edges, each of cost at most $d_U$.}
\label{fig:local}
\end{figure}
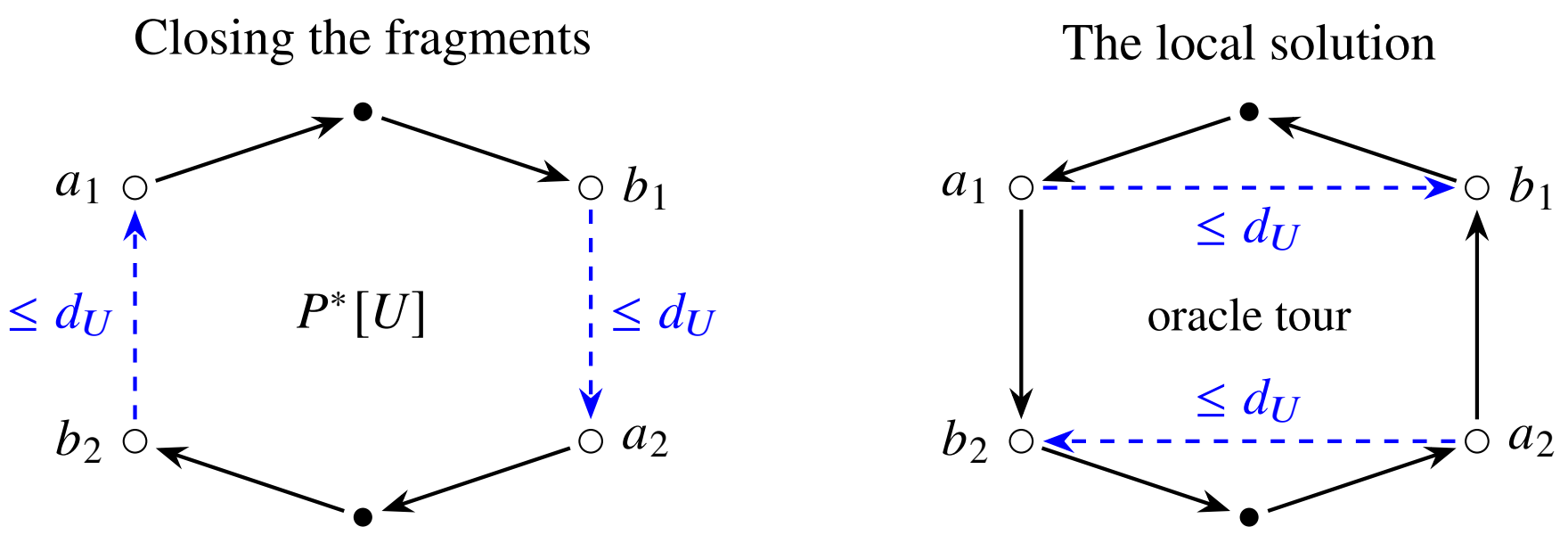

The replacement walks are built like the $s$-$t$ walk above, with the fragments in place of $\Ps$ (Figure~\ref{fig:local}). Closing $\Ps$ by the edge $(t,s)$ becomes linking the fragments into a closed walk, by an edge from the exit of each fragment to the entry of the next one and from the exit of the last fragment to the entry of the first. This gives a tour of $U$ of cost at most $c(\Ps[U])+r_{\Ps}(U)d_U$, where $\Ps[U]$ is the set of edges of $\Ps$ inside $U$, so the tour of $U$ returned by the oracle, the \emph{oracle tour}, costs at most $\alpha$ times this. Going from $s$ to $t$ becomes taking an edge from each entry to its exit, again at a total cost of at most $r_{\Ps}(U)d_U$. Inserting the oracle tour at the start of one of these edges gives one walk for each fragment, with the same entry and exit, and together these walks visit all of $U$. We call these walks a \emph{local solution} of $U$; it costs at most
\[
 \alpha c(\Ps[U])+(\alpha+1)r_{\Ps}(U)d_U.
\]

\paragraph{Patching the parts.}
Now fix, in the analysis, a partition of $V$ into \emph{parts}. Keep the set $K$ of edges of $\Ps$ between different parts, and replace the fragments of $\Ps$ in each part $U$ by a local solution of $U$. Since the local solutions keep the entry--exit pairs, following $\Ps$ through the replacements gives an $s$-$t$ walk that visits every vertex, the \emph{hybrid solution}. The edges of $K$ are paid for once, whereas the oracle inflates the rest of $\Ps$ by the factor $\alpha$, so the hybrid solution costs at most
\[
 \alpha\OPT-(\alpha-1)c(K)+(\alpha+1)\sum_U r_{\Ps}(U)d_U
\]
(Lemma~\ref{lem:hybrid}). We call the last term the \emph{local error}. The hybrid solution costs at most $(\alpha+\eps)\OPT$ if the partition satisfies the \emph{stopping condition}
\begin{equation}\label{eq:allowance-overview}
 (\alpha+1)\sum_U r_{\Ps}(U)d_U\;\le\;D,\qquad\text{where } D:=\eps\OPT+(\alpha-1)c(K)
\end{equation}
is the \emph{allowance}. The analysis below refines the partition until~\eqref{eq:allowance-overview} holds, hence the name stopping condition. Both sides depend on the partition: a finer partition can have parts of smaller diameter, and every edge of $\Ps$ that becomes an edge between parts enlarges the allowance. The hybrid solution itself cannot be computed, however: it depends on the partition and uses the edges of $\Ps$ between the parts, and the algorithm knows neither.

\paragraph{Guessing the edges between parts.}
This is where the dynamic program comes in. The algorithm knows neither the partition nor the entry--exit pairs of $\Ps$, so the dynamic program tries them: it builds walks on vertex sets from smaller to larger ones and, whenever it joins two sets, it guesses the edges between them. It works with a family of polynomially many vertex sets that contains $V$; how the algorithm obtains this family is described below. A \emph{state} consists of a set of the family and a multiset of ordered pairs of its vertices. For each state, the dynamic program computes walks inside the set, one from the first to the second vertex of each pair, that together visit the whole set. For each set, it considers all states with at most $B_*$ pairs, where $B_*$ is a bound that depends only on $\alpha$ and $\eps$ and is established below. So whenever $\Ps$ has at most $B_*$ fragments in a set, the entry--exit pairs of these fragments form one of the states considered, although the algorithm does not know them.

\begin{figure}[htbp]
\centering
\begin{tikzpicture}[scale=0.95,>=Stealth,
  vtx/.style={circle,fill=black,inner sep=1.3pt},
  walk/.style={thick,->,shorten >=2pt,shorten <=2pt},
  guess/.style={blue,very thick,->,shorten >=2pt,shorten <=2pt}]
  \draw[gray] (0,0) ellipse (2.1 and 1.45);
  \draw[gray] (5.4,0) ellipse (2.1 and 1.45);
  \node at (0,1.8) {$U_1$};
  \node at (5.4,1.8) {$U_2$};
  \node[vtx,label=left:$a$] (a) at (-1.4,0.6) {};
  \node[vtx,label=above:$b$] (b) at (1.3,0.65) {};
  \node[vtx,label=below:$c$] (c) at (1.3,-0.65) {};
  \node[vtx,label=left:$d$] (d) at (-1.4,-0.6) {};
  \node[vtx,label=above:$e$] (e) at (4.1,0.65) {};
  \node[vtx,label=below:$f$] (f) at (4.1,-0.65) {};
  \node[vtx,label=above:$g$] (g) at (5.6,0.7) {};
  \node[vtx,label=below:$h$] (h) at (6.6,-0.6) {};
  \draw[walk] (a) .. controls (-0.6,1.3) and (0.4,0.0) .. (b);
  \draw[walk] (c) .. controls (0.4,-1.3) and (-0.6,0.0) .. (d);
  \draw[walk] (e) .. controls (5.1,0.5) and (5.0,-0.5) .. (f);
  \draw[walk] (g) .. controls (6.6,0.9) and (5.7,-0.4) .. (h);
  \draw[guess] (b) -- (e);
  \draw[guess] (f) -- (c);
\end{tikzpicture}
\caption{Joining two sets in the dynamic program. Walks kept for a state on $U_1$ have the pairs $(a,b)$ and $(c,d)$, walks kept for a state on $U_2$ have the pairs $(e,f)$ and $(g,h)$, and the guessed edges between the two sets are $(b,e)$ and $(f,c)$ (blue). The walk from $a$ to $b$, the edge $(b,e)$, the walk from $e$ to $f$, the edge $(f,c)$, and the walk from $c$ to $d$ form one walk from $a$ to $d$; the walk from $g$ to $h$ stays as it is. The result is kept for the state on $U_1\cup U_2$ with the pairs $(a,d)$ and $(g,h)$.}
\label{fig:combine}
\end{figure}
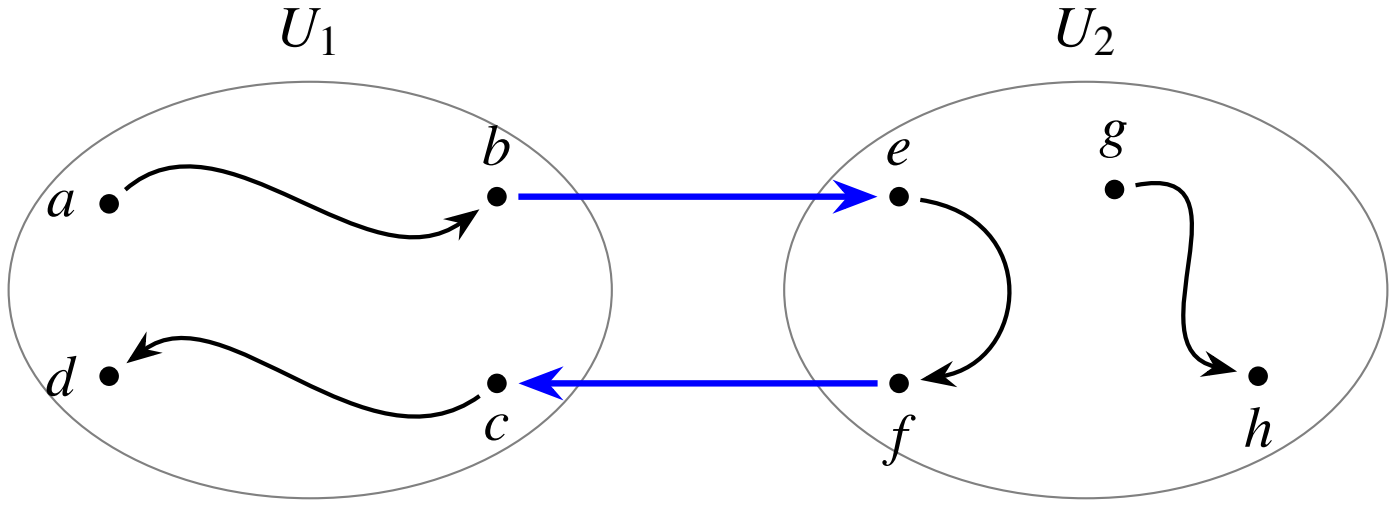

For each state, the dynamic program computes a local solution from one oracle tour of the set; it also divides the set in all possible ways into two sets of the family and joins walks already kept for these two sets by guessed edges between them (Figure~\ref{fig:combine}). For each state, it keeps the cheapest walks it obtains. Its output is the walk kept for $V$ with the single pair $(s,t)$, an $s$-$t$ walk that visits every vertex. The dynamic program of Traub and Vygen~\cite{TV25} for Multi-Path TSP uses states of the same kind; the two differ in how the vertex sets of the states are constructed and in how the walks inside them are computed.

In the analysis, suppose that the partition arises from repeated binary divisions starting with $V$. We record them in the \emph{refinement tree}: each node is a vertex set, the root is $V$, and the two children of a node are the two sets into which it is divided. The leaves are the parts. If every set of the refinement tree belongs to the family and has at most $B_*$ fragments, the dynamic program can follow this tree, and its output costs no more than the hybrid solution for the leaves (Theorem~\ref{thm:dp}). This raises three requirements: (i)~the family must contain every set of the refinement tree, although this tree depends on $\Ps$, and it must have polynomial size; (ii)~the number of fragments must stay bounded in every set of the refinement tree, not just in its leaves, because it determines the number of entry--exit pairs and guessed edges that the dynamic program has to try (Figure~\ref{fig:combine}); (iii)~the leaves must satisfy the stopping condition~\eqref{eq:allowance-overview}. Requirements~(i) and~(ii) keep the dynamic program polynomial, and~(iii) makes its output cost at most $(\alpha+\eps)\OPT$. Requirement~(iii) calls for dividing the sets finely, which conflicts with~(ii), since a division may increase the number of fragments.

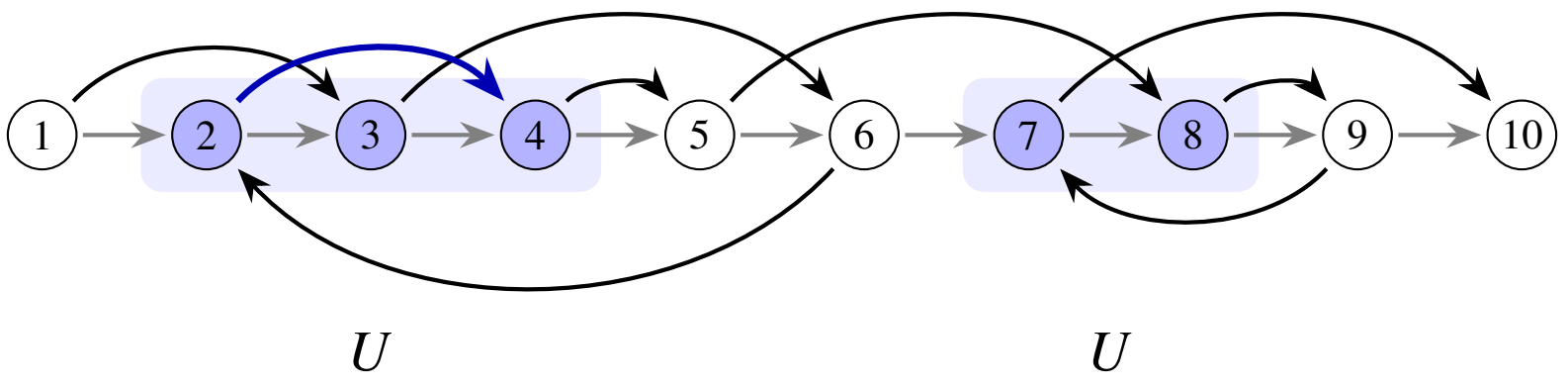
\begin{figure}[htbp]
\centering
\begin{tikzpicture}[xscale=1.1,>=Stealth,
  v/.style={circle,draw,minimum size=0.46cm,inner sep=0pt,font=\scriptsize},
  vin/.style={v,fill=blue!30},
  qe/.style={->,gray,thick,shorten >=1pt,shorten <=1pt},
  pe/.style={->,thick,shorten >=2pt,shorten <=2pt}]
  \fill[blue!8,rounded corners] (1.6,-0.38) rectangle (4.4,0.38);
  \fill[blue!8,rounded corners] (6.6,-0.38) rectangle (8.4,0.38);
  \foreach \i/\st in {1/v,2/vin,3/vin,4/vin,5/v,6/v,7/vin,8/vin,9/v,10/v}{
    \node[\st] (v\i) at (\i,0) {\i};
  }
  \foreach \a/\b in {1/2,2/3,3/4,4/5,5/6,6/7,7/8,8/9,9/10}{\draw[qe] (v\a)--(v\b);}
  \foreach \a/\b in {1/3,3/6,6/2,4/5,5/8,8/9,9/7,7/10}{\draw[pe] (v\a) to[bend left=50] (v\b);}
  \draw[pe,blue!70!black,very thick] (v2) to[bend left=50] (v4);
  \node at (3,-1.45) {$U$};
  \node at (7.5,-1.45) {$U$};
\end{tikzpicture}
\caption{Fragments and intervals of $Q$. The path $Q$ (gray) visits the vertices from left to right; the optimum path $\Ps$ (curved edges) visits them in the order $1,3,6,2,4,5,8,9,7,10$. In $U=\{2,3,4,7,8\}$ (shaded), $\Ps$ has four fragments: $3$, the walk from $2$ to $4$ (blue edge), $8$, and $7$. The path $Q$ has two fragments in $U$, so $U$ is the union of two intervals of $Q$.}
\label{fig:qorder}
\end{figure}

\paragraph{A second reference path.}
A family that satisfies~(i) comes from a second path, which the algorithm does know. The algorithm begins by computing an $s$-$t$ path $Q$ through all vertices of cost at most $\beta\OPT$ for a constant $\beta$; the Feige--Singh reduction gives $\beta=3\alpha$ and uses only the oracle. Fragments of $Q$ are defined like those of $\Ps$. Call a set of vertices that $Q$ visits consecutively an \emph{interval} of $Q$. A set with few fragments of $Q$ is a union of few intervals of $Q$; see Figure~\ref{fig:qorder}. The dynamic program considers all unions of at most $B_*$ such intervals. This family does not depend on $\Ps$, it has only polynomially many members, and it can be enumerated in polynomial time. It contains every set of the refinement tree, provided that these sets also have few fragments of $Q$. So~(ii) now applies to $Q$ as well, and from now on we count the fragments of $\Ps$ and $Q$ together. Together, $\Ps$ and $Q$ cost at most $(1+\beta)\OPT$.

\paragraph{Dividing along levels.}
It remains to construct, in the analysis, a refinement tree that satisfies~(ii) and~(iii). Since requirement~(iii) involves the diameters of the parts, the analysis divides a part along distances from one of its vertices. Starting from $V$, as long as the current partition fails the stopping condition, it repeatedly processes a part $U$ with $d_U>0$; a part $U$ with $d_U=0$, such as a single vertex, adds nothing to the local error and stays a leaf. To process $U$, choose an ordered pair $(a,b)$ with $c(a,b)=d_U$ and give every vertex $v$ of $U$ the height $f(v)=c(a,v)$, a number in $[0,d_U]$. A \emph{level} $\theta\in[0,d_U)$ divides $U$ into two children, the vertices of height at most $\theta$ and those of height above $\theta$. Since $c(a,b)=d_U$, the heights run from $0$ at $a$ to $d_U$ at $b$, so every level separates $a$ from $b$, and a part with a large diameter offers a long range of levels to divide at.

To see how many fragments such a division creates, say that an edge $(u,v)$ inside $U$ \emph{crosses $\theta$ upward} if $f(u)\le\theta<f(v)$, and call the number of edges of $\Ps$ and $Q$ inside $U$ that do so the \emph{upward crossing number} $\up_U(\theta)$ of $\theta$. Between two consecutive visits of a walk to the same child, the walk crosses $\theta$ upward: an excursion into the upper child begins with an upward crossing, and an excursion into the lower child ends with one. So each of the two children, on its own, has at most $\up_U(\theta)$ more fragments than $U$; downward crossings need not be counted separately.

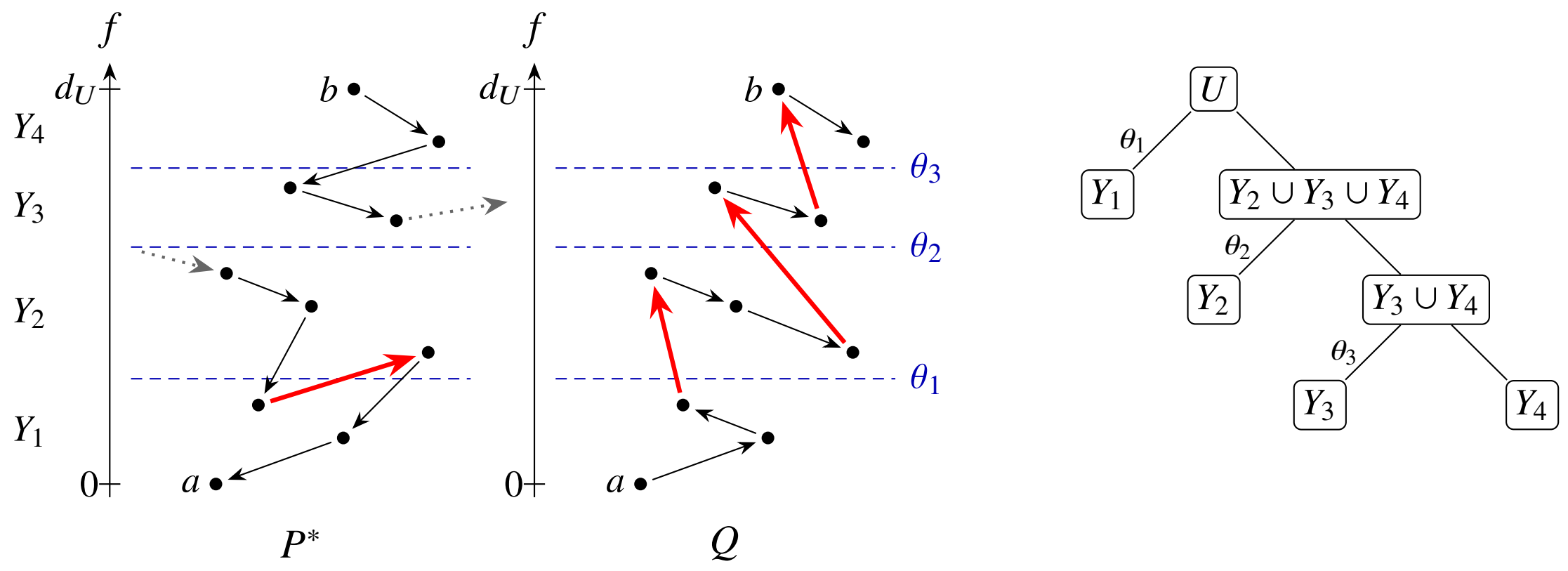
\begin{figure}[htbp]
\centering
\begin{tikzpicture}[>=Stealth,yscale=0.62,
  vtx/.style={circle,fill=black,inner sep=1.2pt},
  tn/.style={draw,rounded corners=2pt,inner sep=2.5pt,font=\small}]
  \foreach \dx/\nm in {0/P,4.0/Q}{
    \begin{scope}[xshift=\dx cm]
      \draw[->] (-0.4,-0.2) -- (-0.4,6.4) node[above] {$f$};
      \draw (-0.5,0) -- (-0.3,0) node[left=2pt] {\small $0$};
      \draw (-0.5,6) -- (-0.3,6) node[left=2pt] {\small $d_U$};
      \foreach \y in {1.6,3.6,4.8}{\draw[blue!70!black,densely dashed] (-0.2,\y) -- (3.0,\y);}
      \foreach \n/\x/\y in {\nm0/0.6/0,\nm1/1.8/0.7,\nm2/1.0/1.2,\nm3/2.6/2.0,\nm4/1.5/2.7,\nm5/0.7/3.2,\nm6/2.3/4.0,\nm7/1.3/4.5,\nm8/2.7/5.2,\nm9/1.9/6}{
        \node[vtx] (\n) at (\x,\y) {};
      }
    \end{scope}
  }
  \node[left] at (P0) {\small $a$}; \node[left] at (P9) {\small $b$};
  \node[left] at (Q0) {\small $a$}; \node[left] at (Q9) {\small $b$};
  \foreach \y/\l in {1.6/1,3.6/2,4.8/3}{\node[blue!70!black,right] at (7.0,\y) {\small $\theta_\l$};}
  \foreach \y/\l in {0.8/1,2.6/2,4.2/3,5.4/4}{\node at (-1.15,\y) {\small $Y_\l$};}
  \begin{scope}[shorten >=1.5pt,shorten <=1.5pt]
    \foreach \u/\w in {P9/P8,P8/P7,P7/P6,P5/P4,P4/P2,P3/P1,P1/P0}{\draw[->] (\u)--(\w);}
    \draw[->,black!60,dotted,thick] (P6) -- (3.4,4.3);
    \draw[->,black!60,dotted,thick] (-0.15,3.55) -- (P5);
    \draw[->,red,very thick] (P2)--(P3);
    \foreach \u/\w in {Q0/Q1,Q1/Q2,Q5/Q4,Q4/Q3,Q7/Q6,Q9/Q8}{\draw[->] (\u)--(\w);}
    \foreach \u/\w in {Q2/Q5,Q3/Q7,Q6/Q9}{\draw[->,red,very thick] (\u)--(\w);}
  \end{scope}
  \node at (1.4,-0.9) {$\Ps$};
  \node at (5.4,-0.9) {$Q$};
  \node[tn] (r) at (10.0,6.0) {$U$};
  \node[tn] (y1) at (9.0,4.4) {$Y_1$};
  \node[tn] (u2) at (11.0,4.4) {$Y_2\cup Y_3\cup Y_4$};
  \node[tn] (y2) at (10.0,2.8) {$Y_2$};
  \node[tn] (u3) at (12.0,2.8) {$Y_3\cup Y_4$};
  \node[tn] (y3) at (11.0,1.2) {$Y_3$};
  \node[tn] (y4) at (13.0,1.2) {$Y_4$};
  \draw (r) -- node[left,font=\scriptsize] {$\theta_1$} (y1); \draw (r) -- (u2);
  \draw (u2) -- node[left,font=\scriptsize] {$\theta_2$} (y2); \draw (u2) -- (u3);
  \draw (u3) -- node[left,font=\scriptsize] {$\theta_3$} (y3); \draw (u3) -- (y4);
\end{tikzpicture}
\caption{Processing a part $U$ at three levels $\theta_1<\theta_2<\theta_3$ of its height function $f$, which cut $U$ into $Y_1,\ldots,Y_4$ (left). Left and middle: the vertices of $U$ at their heights, with the two fragments of $\Ps$ in $U$ (dotted arrows show where $\Ps$ leaves and reenters $U$) and a fragment of $Q$. The chosen levels are dashed, and the edges that cross a chosen level upward are red. The level $\theta_1$ is crossed upward by both paths, and $\theta_2$ and $\theta_3$ only by $Q$, so the upward crossing numbers are $2$, $1$ and $1$. Right: the refinement tree first divides $U$ at $\theta_1$, then the upper set at $\theta_2$, then at $\theta_3$. Every set created is bounded by one or two of these levels, such as $Y_2$ by $\theta_1$ and $\theta_2$, or $Y_2\cup Y_3\cup Y_4$ by $\theta_1$ alone, so it is a slab of $U$ and the slab bound applies. For example, $\Ps$ and $Q$ have three fragments in $U$ and three upward crossings of the boundary levels $\theta_1,\theta_2$ of $Y_2$, so $Y_2$ has at most six fragments of them; it has three.}
\label{fig:division}
\end{figure}

The analysis may divide $U$ at several levels $\theta_1<\cdots<\theta_m$ of its height function $f$ at once. It records this as a sequence of binary divisions: first at $\theta_1$, then the upper of the two sets at $\theta_2$, and so on (Figure~\ref{fig:division}); the levels are chosen so that every set that arises is nonempty. Every set that arises consists of the vertices of $U$ whose heights lie between two chosen levels, or on one side of one chosen level; we call such a set a \emph{slab} of $U$, and the one or two chosen levels that bound it its \emph{boundary levels}. By the argument above, a walk that leaves a slab and comes back crosses one of the boundary levels of the slab upward, and the chosen levels inside the slab do not matter. So the number of fragments of $\Ps$ and $Q$ in a slab is at most their number in $U$ plus the upward crossing numbers of the boundary levels of the slab, however many levels are chosen (Figure~\ref{fig:division} and Lemma~\ref{lem:fragments}). This \emph{slab bound} is what allows dividing at many levels at once: every set created is affected only by its own boundary levels. The remaining question is which levels to choose, so that the upward crossing numbers of the boundary levels stay under control. There are two ways.

\begin{figure}[H]
\centering
\begin{tikzpicture}[xscale=0.95,yscale=0.6,>=Stealth,
  vtx/.style={circle,fill=black,inner sep=1.4pt},
  refe/.style={thick,->,shorten >=1.5pt,shorten <=1.5pt}]
  \fill[green!12] (-0.4,1.9) rectangle (8.2,4.25);
  \draw[->] (-1.0,-0.3) -- (-1.0,5.6) node[above] {$f$};
  \draw (-1.1,0) -- (-0.9,0) node[left=3pt] {$0$};
  \draw (-1.1,5) -- (-0.9,5) node[left=3pt] {$c(t,s)$};
  \node[vtx,label=left:$s$] (s) at (0.2,5.0) {};
  \node[vtx] (a) at (1.2,4.3) {};
  \node[vtx] (b) at (2.0,4.6) {};
  \node[vtx] (c) at (3.0,3.8) {};
  \node[vtx] (d) at (3.8,3.4) {};
  \node[vtx] (e) at (4.6,1.6) {};
  \node[vtx] (f) at (5.6,1.85) {};
  \node[vtx] (g) at (6.6,0.8) {};
  \node[vtx,label=below:$t$] (t) at (7.6,0.0) {};
  \draw[refe] (s)--(a); \draw[refe,red] (a)--(b); \draw[refe] (b)--(c); \draw[refe] (c)--(d);
  \draw[refe] (d)--(e); \draw[refe,red] (e)--(f); \draw[refe] (f)--(g); \draw[refe] (g)--(t);
  \draw[dashed,->,blue,thick] (t) .. controls (9.2,2.5) and (5.5,6.2) .. node[pos=0.12,right] {$d_V=c(t,s)$} (s);
  \draw[red,densely dashed,very thick] (-0.4,2.6) -- (8.2,2.6);
  \node[red,above] at (1.4,2.62) {free split};
\end{tikzpicture}
\caption{A large return distance. The height $f(v)=c(t,v)$ increases from $t$ (bottom) to $s$ (top), and every level between $0$ and $c(t,s)$ separates $t$ from $s$. The blue dashed arrow marks the pair $(t,s)$ attaining the diameter; it is not an edge of a solution. The optimum path $\Ps$ (black) mostly goes down, and its upward edges (red) cover only a short range of levels. The red dashed level, in the shaded band, has no upward crossing of $\Ps$ or $Q$ ($Q$ is not drawn), so each path has a single fragment on each side of this free split.}
\label{fig:return}
\end{figure}
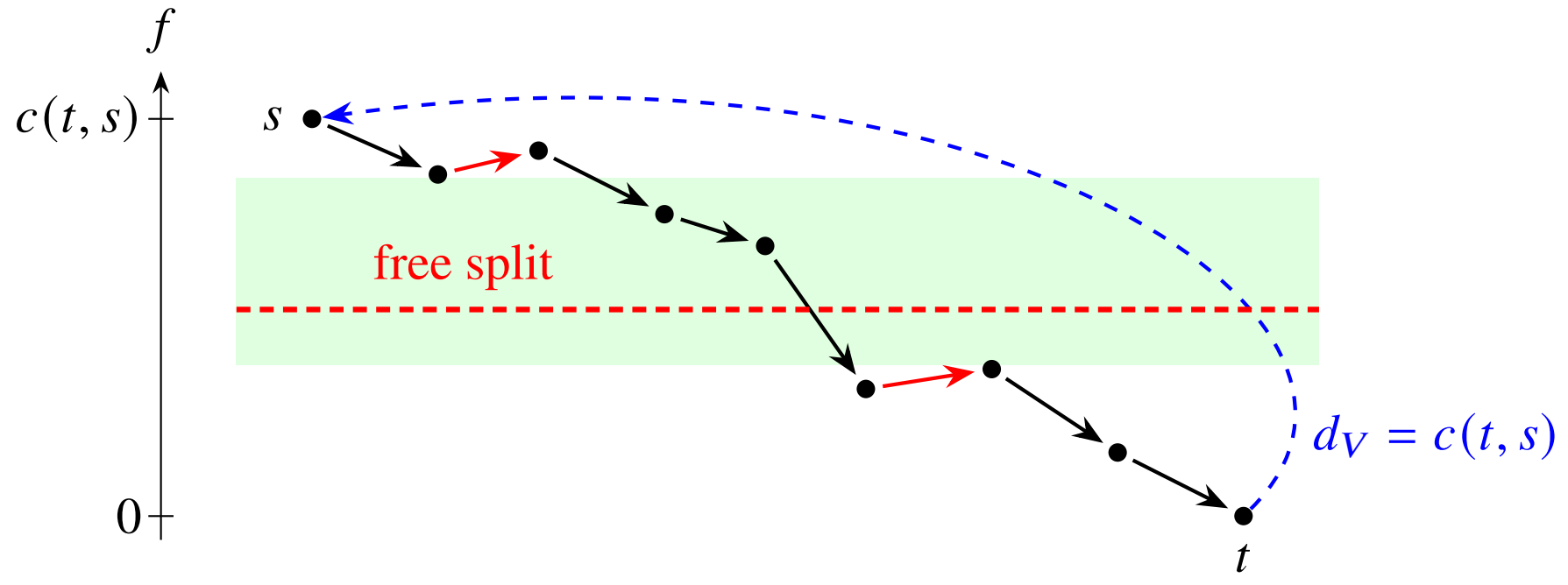

\paragraph{Free splits.}
The ideal level is one that no edge of $\Ps$ or $Q$ crosses upward. If some level of a part $U$ has upward crossing number zero, the analysis processes $U$ with this level alone. By the slab bound, neither child has more fragments than $U$. We call such a division a \emph{free split}. Such levels exist whenever $U$ has a large diameter. To see this, we measure sets of levels on the height axis: the \emph{measure} $\meas\Theta$ of a set $\Theta$ of levels that is a finite union of intervals is the total length of these intervals, and the levels of $U$ form the interval $[0,d_U)$ of measure $d_U$. An edge $(u,v)$ with $f(u)<f(v)$ crosses upward exactly the levels in $[f(u),f(v))$, a set of measure $f(v)-f(u)\le c(u,v)$ by the triangle inequality; other edges cross no level upward. So the levels crossed upward by some edge of $\Ps$ or $Q$ have measure at most $c(\Ps)+c(Q)\le(1+\beta)\OPT$, and every part $U$ with $d_U>(1+\beta)\OPT$ has a free split. In particular, if the return distance $c(t,s)$, which Traub, Vygen and Zenklusen identified as a difficulty, exceeds $(1+\beta)\OPT$, then so does $d_V$, and $V$ has a free split (Figure~\ref{fig:return}). In $V$, the path $\Ps$ crosses such a level exactly once, so the split amounts to guessing a single edge of $\Ps$.

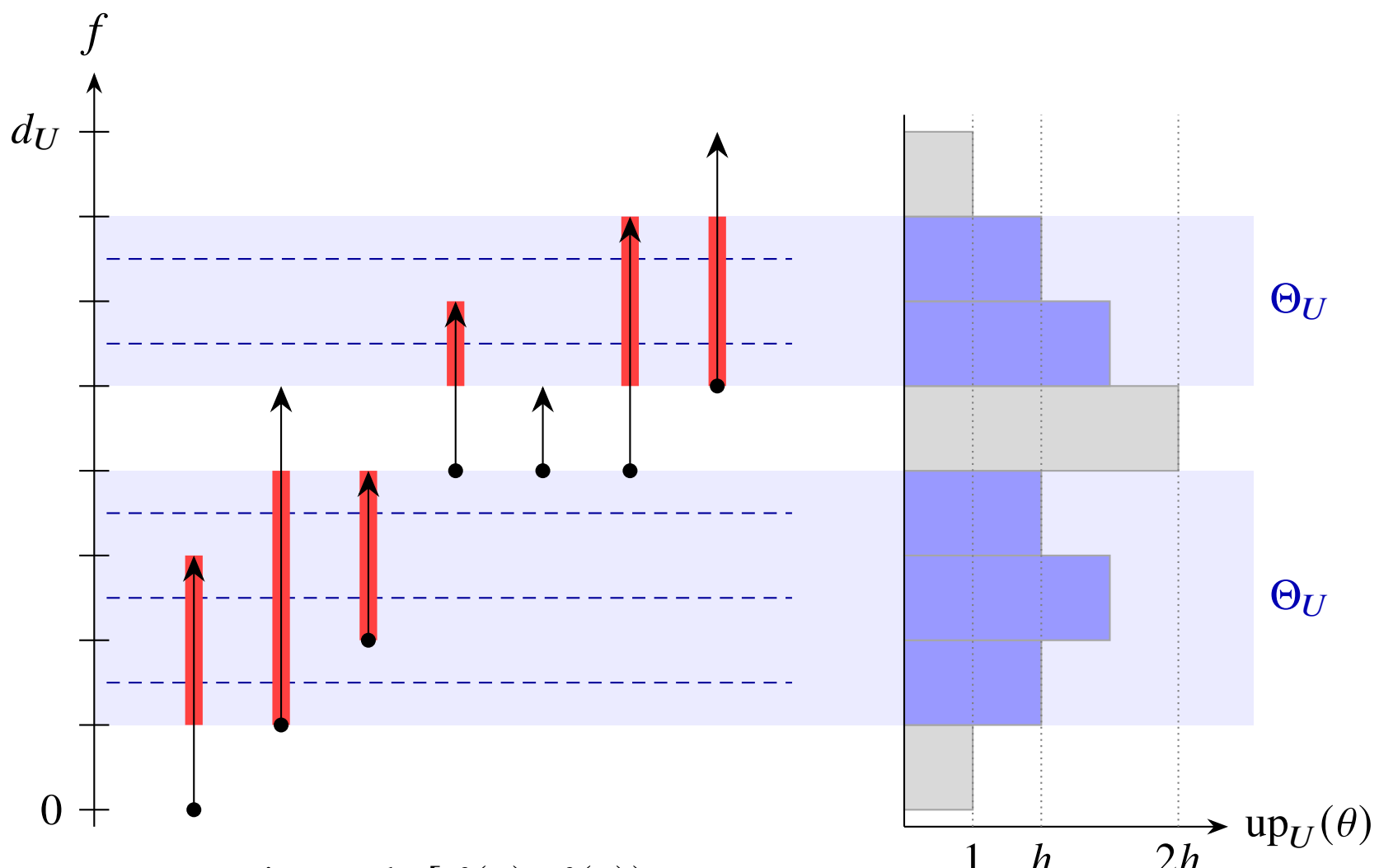
\begin{figure}[H]
\centering
\begin{tikzpicture}[yscale=0.68,
  vtx/.style={circle,fill=black,inner sep=1.2pt},
  edge/.style={-{Stealth[length=6pt,width=5pt]},black,thin}]
  \foreach \lo/\hi in {1/4,5/7}{\fill[blue!8] (-0.3,\lo) rectangle (9.0,\hi);}
  \node[blue!70!black,right] at (9.0,2.5) {\small $\Theta_U$};
  \node[blue!70!black,right] at (9.0,6) {\small $\Theta_U$};
  \draw[-{Stealth}] (-0.3,-0.2) -- (-0.3,8.7) node[above] {$f$};
  \foreach \y in {0,...,8}{\draw (-0.42,\y) -- (-0.18,\y);}
  \node[left] at (-0.42,0) {\small $0$};
  \node[left] at (-0.42,8) {\small $d_U$};
  \foreach \y in {1.5,2.5,3.5,5.5,6.5}{\draw[blue!60!black,densely dashed] (-0.2,\y) -- (5.3,\y);}
  \foreach \x/\lo/\hi in {0.5/1/3,1.2/1/4,1.9/2/4,2.6/5/6,4.0/5/7,4.7/5/7}{
    \draw[red!75,line width=4pt] (\x,\lo) -- (\x,\hi);
  }
  \foreach \x/\lo/\hi in {0.5/0/3,1.2/1/5,1.9/2/4,2.6/4/6,3.3/4/5,4.0/4/7,4.7/5/8}{
    \draw[edge] (\x,\lo) -- (\x,\hi);
    \node[vtx] at (\x,\lo) {};
  }
  \node at (2.6,-0.9) {\small intervals $[f(u),f(v))$};
  \foreach \lo/\hi/\u/\sel in {0/1/1/0,1/2/2/1,2/3/3/1,3/4/2/1,4/5/4/0,5/6/3/1,6/7/2/1,7/8/1/0}{
    \ifnum\sel=1 \fill[blue!40] (6.2,\lo) rectangle ({6.2+0.55*\u},\hi);
    \else \fill[gray!30] (6.2,\lo) rectangle ({6.2+0.55*\u},\hi); \fi
    \draw[gray!70] (6.2,\lo) rectangle ({6.2+0.55*\u},\hi);
  }
  \draw (6.2,-0.2) -- (6.2,8.2);
  \draw[-{Stealth}] (6.2,-0.2) -- (8.8,-0.2) node[right] {\small $\up_U(\theta)$};
  \foreach \u/\l in {1/1,2/h,4/2h}{
    \draw[densely dotted,gray] ({6.2+0.55*\u},-0.2) -- ({6.2+0.55*\u},8.2);
    \node[below] at ({6.2+0.55*\u},-0.2) {\small $\l$};
  }
\end{tikzpicture}
\caption{Processing a part $U$ without a free split, for $h=2$. Left: every edge $(u,v)$ of $\Ps$ or $Q$ inside $U$ with $f(u)<f(v)$ is drawn, at its own horizontal position, as the interval $[f(u),f(v))$ of the levels that it crosses upward, from $f(u)$ (dot) to $f(v)$ (arrowhead); edges that do not go up cross no level upward and are omitted. Right: the upward crossing number $\up_U(\theta)$ of every level $\theta$, the number of intervals on the left that contain $\theta$. The ticks on the axis are the vertex heights. The levels with upward crossing number in $[h,2h)$ form the shaded set $\Theta_U$; it consists of five intervals between consecutive ticks, which form two shaded bands, and $U$ is divided at one level in each of the five (dashed). Every edge whose interval meets $\Theta_U$ crosses a dashed level upward and is cut. The part of its interval in $\Theta_U$ (red) has measure at most $f(v)-f(u)\le c(u,v)$. Each level of $\Theta_U$ lies in at least $h$ red pieces, so the red pieces have total measure $\int_{\Theta_U}\up_U(\theta)\,d\theta\ge h\meas\Theta_U$, and the cut edges cost at least this much. The only interval without a red piece lies outside $\Theta_U$, and its edge is not cut.}
\label{fig:coarea}
\end{figure}

\paragraph{Paying for the divisions.}
If no part has a free split, every level of every part has a positive upward crossing number, and a division may create new fragments. Suppose that every part $U$ of the current partition has at most $B$ fragments of $\Ps$ and $Q$ in total. We call the edges of $\Ps$ and $Q$ that the processing of $U$ turns into edges between parts the \emph{cut edges} of $U$. The analysis bounds the growth of the number of fragments by the cost of the cut edges, as explained below. A single level does not suffice for this: the edges cut at one level may all be cheap, while each child may get as many new fragments as the upward crossing number of the level. The analysis therefore divides each part at many levels at once, all with similar upward crossing numbers. It processes all parts in one step. Fix $h\ge1$, the same for all parts, and let $\Theta_U$ be the set of levels of a part $U$ whose upward crossing number lies in $[h,2h)$. All levels between two consecutive vertex heights give the same division of $U$ and have the same upward crossing number, so $\Theta_U$ is a union of such intervals. The analysis divides $U$ at one level from each interval between consecutive vertex heights that lies in $\Theta_U$ (Figure~\ref{fig:coarea}).

Dividing at many levels does not multiply the new fragments. Each set created from $U$ has at most two boundary levels, each of upward crossing number less than $2h$. So, by the slab bound, every such set has fewer than $B+4h$ fragments, however many levels are used. The cost of the cut edges behaves differently: Figure~\ref{fig:coarea} explains why the multiset $E_U$ of cut edges of $U$ satisfies
\begin{equation}\label{eq:coarea-overview}
 c(E_U)\;\ge\;\int_{\Theta_U}\up_U(\theta)\,d\theta\;\ge\;h\meas\Theta_U .
\end{equation}
This contrast is the key point of the construction: the growth of the number of fragments in each new set is governed by at most two levels, its boundary levels, whereas the cost of the cut edges accumulates over all chosen levels. Since the chosen levels have similar upward crossing numbers, the two sides are comparable: an increase of $4h$ in the bound $B$, against a cost of at least $h$ per unit of measure of $\Theta_U$. Both are proportional to $h$, so what decides is the total measure $\sum_U\meas\Theta_U$, not $h$: once it is large, the increase of $B$ is paid for by the cost of the cut edges, which is at least $h\sum_U\meas\Theta_U$. Any fixed ratio between the largest and the smallest upward crossing numbers of the chosen levels would do; the factor $2$ in $[h,2h)$ is chosen for simplicity.

It remains to show that some $h$ gives sets $\Theta_U$ of large total measure, and here the stopping condition enters. Since every part $U$ has $r_{\Ps}(U)\le B$, failing the stopping condition~\eqref{eq:allowance-overview} implies $\sum_Ud_U>D/((\alpha+1)B)$: the levels of all parts have total measure more than $D/((\alpha+1)B)$. Few of these levels can have a very large upward crossing number, because the cost of $\Ps$ and $Q$ limits how often the levels are crossed upward on average. Indeed, by the argument of Figure~\ref{fig:coarea}, applied to all levels, the integral of the upward crossing numbers over all levels of all parts is at most the cost of $\Ps$ and $Q$, that is, at most $(1+\beta)\OPT$. By the definition of $D$ in~\eqref{eq:allowance-overview}, $D\ge\eps\OPT$, so this integral is at most $(1+\beta)D/\eps$. Consequently, the levels whose upward crossing number exceeds a large constant times $B$ have total measure at most $(1+\beta)D/\eps$ divided by this multiple of $B$, only a small fraction of $D/((\alpha+1)B)$. The remaining levels of all parts fall into logarithmically many ranges $[h,2h)$. Adding up their measures over all parts, some range has at least the average total, and the analysis takes its $h$ for all parts. Then $\sum_U\meas\Theta_U$ is at least $D/((\alpha+1)B)$ up to a constant and a factor logarithmic in $B$ (Section~\ref{sec:refinement-step}), although a single part may have few levels in this range. For this $h$, \eqref{eq:coarea-overview} shows that the cut edges cost at least $(4h/B)\cdot D$, up to a constant and a factor logarithmic in $B$ (Lemma~\ref{lem:refinement-step}): compared with the allowance $D$, the cost of the cut edges matches the relative growth $4h/B$ of the bound on the number of fragments.

\paragraph{Adding up the payments.}
So far, $B$ was the bound on the number of fragments at the current step. Each step replaces it by $B+4h$, or keeps it after a free split. By the previous paragraph, the relative growth $4h/B$ is at most the cost of the cut edges relative to $D$, times a constant and a factor logarithmic in $B$; this factor comes from the logarithmically many ranges $[h,2h)$. Looking at $\log B$ instead of $B$ removes this factor. Since the increase of $\log B$ is at most the relative increase of $B$ (taking natural logarithms), the relative increase of $\log B$ is at most that of $B$ divided by $\log B$, and this division cancels the factor logarithmic in $B$. So in each step, the relative growth of $\log B$ is at most a constant times the cost of the cut edges relative to $D$ (Lemma~\ref{lem:logstep}). An edge that has been cut joins two different parts from then on, so it is never cut again. All steps together therefore cut edges of total cost at most $(1+\beta)\OPT$, the cost of $\Ps$ and $Q$, while $D\ge\eps\OPT$ at every step. So the relative growths of $\log B$ over all steps add up to at most a constant times $(1+\beta)/\eps$, and $\log B$, and with it $B$, stays below a bound $B_*$ that depends only on $\alpha$ and $\eps$. This bound holds at every step of the refinement, so every set that ever arises in the refinement tree has fewer than $B_*$ fragments of $\Ps$ and $Q$, which gives~(ii). Finally, the analysis stops as soon as the stopping condition holds, which gives~(iii). Every step divides at least one part, so this happens after at most $n-1$ steps, at the latest when all parts are single vertices, whose diameters are zero.

This accounting already bounds $B$ by a constant, which is all the dynamic program needs; the constant is doubly exponential in $1/\eps$. The proof sharpens it to a bound that is singly exponential in a polynomial in $1/\eps$ for fixed $\alpha$, using that the cut edges of $\Ps$ also enlarge $D$ (Section~\ref{sec:width}). For this, it weights the fragments of $\Ps$ by a large constant, so that the cut edges of $Q$, which do not enlarge $D$, have only a small effect (Section~\ref{sec:structure}); this overview counts fragments of $\Ps$ and $Q$ alike.

\paragraph{Why the dynamic program stays polynomial.}
The bound $B_*$ depends only on $\alpha$ and $\eps$, so the algorithm can compute it in advance, without knowing the refinement tree. The family consists of the unions of at most $B_*$ intervals of $Q$, and every state carries at most $B_*$ entry--exit pairs, so for fixed $\alpha$ and $\eps$ there are polynomially many states and transitions. The divisions take place only in the analysis: the algorithm computes $Q$, fills one table, and shortcuts the resulting $s$-$t$ walk (Algorithm~\ref{alg:main} in Section~\ref{sec:dp}).

\section{The setting and fragments}\label{sec:prelim}

\subsection{Directed metrics and walks}

Let $V$ be a set of $n$ vertices, and let $c$ be a directed metric on $V$: a function $c\colon V\times V\to\mathbb Q_{\ge0}$ with $c(v,v)=0$ and $c(u,w)\le c(u,v)+c(v,w)$ for all $u,v,w\in V$, which need not be symmetric. As in Section~\ref{sec:overview}, we regard $c$ as the complete directed graph on $V$ in which the edge $e=(u,v)$, for $u\ne v$, costs $c(e)=c(u,v)$. For a multiset $F$ of edges, $c(F)$ is its total cost.

A \emph{walk} is a finite vertex-and-edge sequence in this graph. It may repeat vertices and edges; we speak of the occurrences of a vertex and the uses of an edge in the walk, and the cost of the walk counts every edge use. A walk consisting of one vertex, a \emph{zero-length walk}, has cost zero. A walk is \emph{closed} if its first and last vertices agree, and \emph{open} otherwise. We regard a closed walk as a linear sequence with a specified first vertex, so that closed and open walks can be treated alike below. A \emph{path} is a walk that repeats no vertex, and a \emph{tour} is a closed walk that repeats no vertex apart from its return to the first one. For a nonempty set $U$, a \emph{tour of $U$} is a tour whose vertices are exactly those of $U$.

By the triangle inequality, shortcutting repeated occurrences of vertices does not increase the cost. It turns every closed walk whose vertices are exactly those of $U$ into a tour of $U$, and every walk between two distinct vertices into a path between them with the same vertices. Together with taking shortest-path distances, this shows that Asymmetric TSP and Asymmetric Path TSP are equivalent to their versions on directed graphs with arbitrary nonnegative edge costs, in which vertices may be visited more than once. For Asymmetric Path TSP, the directed graph need not be strongly connected; then pairs of vertices without a connecting path get a sufficiently large distance, so that the triangle inequality still holds and, if the original instance is feasible, no solution within the approximation ratio in question uses such a pair.

\subsection{Several prescribed pairs and the reference family}\label{sec:fragments-def}\label{sec:interface}

A walk may enter and leave a vertex set $U$ several times, and its maximal parts inside $U$ are its fragments. Formally, a \emph{fragment} of a walk $P$ in $U$ is a maximal consecutive subsequence of $P$ all of whose vertices belong to $U$. A fragment may be a single vertex. Different fragments of a walk are counted separately, even if they have the same end vertices or share vertices. The first and last vertices of a fragment are its \emph{entry} and \emph{exit}, and we keep them paired. Write $P[U]$ for the multiset of edge uses of $P$ with both endpoints in $U$; these are exactly the edge uses within its fragments. For a family $\cP$ of walks, the edge multisets and the fragments are taken over all its walks, and $r_{\cP}(U)$ denotes the total number of fragments of its walks in $U$; for a single walk, as in Section~\ref{sec:overview}, this is the number of its fragments. All edge counts distinguish separate uses of the same edge, both within a walk and between walks.

In Section~\ref{sec:overview}, the dynamic program guesses vertex sets and, for each of them, the entry--exit pairs of the fragments of the optimum path in the set; inside the set, it builds one walk for each guessed pair from the oracle tour of the set. Although Section~\ref{sec:overview} considers a single prescribed pair, the dynamic program thus already deals with several entry--exit pairs in each vertex set. We therefore formulate the algorithm and its analysis for an arbitrary fixed number of prescribed pairs, that is, for Asymmetric Multi-Path TSP. The input specifies $k$ ordered pairs $(s_1,t_1),\ldots,(s_k,t_k)$, where $k\ge1$, and we denote their multiset by $M_0$. The task is to find $k$ paths that together visit all of $V$, the $i$th from $s_i$ to $t_i$, or a tour containing $s_i$ if $s_i=t_i$, minimizing their total cost. End vertices may coincide, different pairs may be equal, and the paths may share vertices. We work with walks instead: $k$ walks, the $i$th from $s_i$ to $t_i$, that together visit all of $V$ form a \emph{feasible family}. Every solution is a feasible family, and shortcutting every walk of a feasible family gives a solution of no greater cost, so both have the same minimum cost, which we denote by $\OPT$. The one-vertex case is immediate, so we assume $n\ge2$.

Fix an optimum feasible family $\cPs$; for Asymmetric Path TSP, that is, for one prescribed pair $(s,t)$ with $s\ne t$, we may take it to consist of the optimum path $\Ps$ of Section~\ref{sec:overview}. The algorithm does not know $\cPs$, which is used only in the analysis. As in Section~\ref{sec:overview}, the analysis fixes a partition of $V$ into parts, keeps the edge uses of $\cPs$ between different parts, and replaces the fragments of $\cPs$ in each part by walks inside the part that have the same entry--exit pairs and together visit the part. Following the walks of $\cPs$ through these replacements gives a feasible family, the \emph{hybrid solution} of the partition; Section~\ref{sec:local} builds the replacement walks from oracle tours and bounds the cost of the hybrid solution. The replacement depends on the fragments only through their entry--exit pairs. For a nonempty set $U$, we have $r_{\cPs}(U)\ge1$, since $\cPs$ covers $V$, and we write $M_U$ for the multiset of entry--exit pairs of the fragments of $\cPs$ in $U$; it has $r_{\cPs}(U)$ pairs.

The dynamic program handles several prescribed pairs in the same way as one; the reference path is replaced by a reference family. In Section~\ref{sec:overview}, the vertex sets considered are the unions of few intervals of $Q$, where $Q$ is a feasible solution of Asymmetric Path TSP computed by the Feige--Singh reduction. In general, a family $\cQ$ of vertex-disjoint paths that together visit every vertex, the \emph{reference family}, takes the place of $Q$; for Asymmetric Path TSP, $\cQ$ may consist of the single path $Q$. Section~\ref{sec:overview} explained why it is needed; in the proof, it plays two roles. First, it makes the sets of the analysis enumerable: a set with few fragments of $\cQ$ is a union of few intervals of the order in which $\cQ$ visits the vertices (Lemma~\ref{lem:first-order} below). Since the algorithm knows $\cQ$, the dynamic program can then enumerate every set the analysis uses, as long as the analysis controls the numbers of fragments. Second, it makes the numbers of fragments controllable: their growth during the refinement is paid for by the cost of newly cut edges of $\cPs$ and $\cQ$, and the total cost of $\cQ$ is at most a constant times $\OPT$, so the total payment is bounded. In addition, the refinement starts from $V$, in which every path of $\cQ$ is a single fragment, so the number of paths bounds the initial number of fragments of $\cQ$ and must be a constant as well. This leads to three conditions on the reference family:
\begin{equation}\label{eq:qref}
 V(\cQ)=V,\qquad |\cQ|\le\ell,\qquad c(\cQ)\le\beta\OPT,
\end{equation}
where $V(\cQ)$ is the set of vertices visited by $\cQ$, and $\beta\ge1$ and the positive integer $\ell$ are constants known to the algorithm. The algorithm uses the bound $\beta$, but does not need the value of $\OPT$, and the end vertices of the paths in $\cQ$ need not be the prescribed ones.

The algorithm computes $\cQ$ first. For Asymmetric Path TSP, one run of the Feige--Singh reduction~\cite{FS07} with the oracle and error parameter~$1$ returns an $s$-$t$ path through all vertices of cost at most $3\alpha\OPT$, so~\eqref{eq:qref} holds with $\beta=3\alpha$ and $\ell=1$. Table~\ref{tab:inputs} lists the constants for all problems of this paper; Appendix~\ref{sec:initialization} gives the constructions.

\begin{table}[t]
\centering
\small
\renewcommand{\arraystretch}{1.15}
\begin{tabular}{@{}lcc@{}}
\hline
Problem & $\beta$ & $\ell$\\
\hline
Asymmetric Path TSP & $3\alpha$ & $1$\\
Asymmetric Multi-Path TSP & $\alpha(2k+1)$ & $\lceil\alpha(k+1)\rceil$\\
Asymmetric Multi-Path TSP on unweighted digraphs & $\alpha(2k+1)$ & $\lceil\alpha(k+1)\rceil$\\
Multi-Path TSP, also on unweighted graphs & $2$ & $k$\\
\hline
\end{tabular}
\caption{The constants $\beta$ and $\ell$ of~\eqref{eq:qref} for the reference families used in this paper, with $k$ prescribed pairs. The first family comes from the Feige--Singh reduction; Appendix~\ref{sec:initialization} constructs the others.}
\label{tab:inputs}
\end{table}

Since the paths of $\cQ$ are vertex-disjoint and together visit every vertex, listing their vertices path by path, in a fixed order of the paths, lists every vertex exactly once. This gives an order $\sigma$ of $V$. Call a set of vertices in consecutive positions of $\sigma$ an \emph{interval} of $\sigma$. A fragment of $\cQ$ consists of consecutive vertices of one path, so its vertex set is an interval of $\sigma$. For Asymmetric Path TSP, $\sigma$ is the order in which $Q$ visits the vertices, and its intervals are those of Section~\ref{sec:overview}; see Figure~\ref{fig:qorder}.

\begin{lemma}[intervals]\label{lem:first-order}
Every set $U\subseteq V$ is the union of $r_{\cQ}(U)$ intervals of $\sigma$, one for each fragment of $\cQ$ in $U$.
\end{lemma}

\begin{proof}
Every vertex of $U$ occurs exactly once in the paths of $\cQ$, so it lies in exactly one fragment of $\cQ$ in $U$. Hence the vertex sets of these fragments, which are intervals of $\sigma$, partition $U$.
\end{proof}

The last condition in~\eqref{eq:qref} bounds what is available to pay for the growth of the numbers of fragments: the edges cut during the refinement are edges of $\cPs$ and $\cQ$, and together with $c(\cPs)=\OPT$,
\begin{equation}\label{eq:referencecost}
 c(\cPs)+c(\cQ)\le(1+\beta)\OPT.
\end{equation}

The algorithm then runs the dynamic program on the instance, given by the directed metric $c$ and the multiset $M_0$ of prescribed pairs, with the oracle, the given $\alpha$-approximation algorithm for Asymmetric TSP, and with the reference family $\cQ$, and finally shortcuts the returned walks. The rest of the paper proves that the algorithm runs in polynomial time and that its result costs at most $(\alpha+\eps)\OPT$. Section~\ref{sec:local} builds local solutions from oracle tours and gives a condition on a partition, the stopping condition, under which its hybrid solution costs at most $(\alpha+\eps)\OPT$. Partitions, and their refinement until the stopping condition holds, occur only in the analysis; the algorithm computes no partition. Section~\ref{sec:dp} describes the dynamic program and shows that its output costs no more than the hybrid solution of the leaves of every refinement tree whose sets have at most $B_*$ fragments of the optimum family $\cPs$ and at most $B_*$ fragments of the reference family $\cQ$. Section~\ref{sec:structure} states that such a tree with leaves satisfying the stopping condition exists (Theorem~\ref{thm:process}) and derives Theorems~\ref{thm:main} and~\ref{thm:unified}. Sections~\ref{sec:fragments} to~\ref{sec:width} prove Theorem~\ref{thm:process} by the refinement process outlined in Section~\ref{sec:overview}. Section~\ref{sec:applications} adapts the proof to unweighted digraphs and to the symmetric case, and Appendix~\ref{sec:initialization} constructs the reference families. Section~\ref{sec:overview} explains how the parts fit together.

\section{Local candidates and their cost}\label{sec:local}

In this section we construct local solutions from one oracle tour and bound the cost of the hybrid solution of a partition.

\subsection{Local solutions}\label{sec:local-solutions}

As in Section~\ref{sec:overview}, we want to replace walks inside a vertex set without changing how they enter and leave it. Let $U$ be a nonempty vertex set, and let
\[
 M=\{(a_1,b_1),\ldots,(a_m,b_m)\}
\]
be a nonempty multiset of ordered pairs of vertices of $U$. A \emph{local solution} for $(U,M)$ is a family of $m$ walks inside $U$, that is, with all vertices in $U$, one from $a_i$ to $b_i$ for each $i$, that together visit every vertex of $U$. We may list equal pairs in any fixed order, so two local solutions for $(U,M)$ can be matched walk by walk by their pairs, including multiplicity.

Local solutions for the same $(U,M)$ can replace one another. Suppose that the walks of a local solution for $(U,M)$ are concatenated, together with further edges or walks, into longer walks, each of them used once. Replacing every walk by the walk with the same pair in another local solution for $(U,M)$ keeps every concatenation valid and keeps its end vertices. Both local solutions cover $U$, so coverage is preserved. The total cost changes by the difference of their costs. In particular, the fragments of $\cPs$ in $U$ form a local solution for $(U,M_U)$, where $M_U$ is the multiset of their entry--exit pairs, and replacing them by another one gives again a feasible family.

\subsection{The local candidate}\label{sec:candidate}

Let $U$ be a nonempty set. Run the oracle, the given $\alpha$-approximation algorithm for Asymmetric TSP, once on the restriction of $c$ to $U$, which is again a directed metric. It returns the \emph{oracle tour} $C_U$ of $U$, a tour of $U$ of cost at most $\alpha$ times the minimum cost of a tour of $U$. For a multiset $M$ of pairs as in Section~\ref{sec:local-solutions}, the \emph{connection} of a pair $(a_i,b_i)$ is the edge $(a_i,b_i)$, or the zero-length walk at $a_i$ if $a_i=b_i$. Insert $C_U$ at the start of the first connection: the first walk goes once around $C_U$ from $a_1$ back to $a_1$ and then follows the connection to $b_1$, and every other walk is the connection for its pair. This is a local solution for $(U,M)$, the one built in Section~\ref{sec:overview}; we call it the \emph{local candidate} for $(U,M)$, since the dynamic program also obtains local solutions by combining smaller sets. See Figure~\ref{fig:local}, right. As in Section~\ref{sec:overview}, let
\[
 d_U=\max_{u,v\in U}c(u,v)
\]
be the \emph{diameter} of $U$, taken over ordered pairs. Each connection is an edge between two vertices of $U$ or a zero-length walk, so it costs at most $d_U$, and the local candidate costs at most
\begin{equation}\label{eq:candidatecost}
 c(C_U)+m\,d_U.
\end{equation}
We write $F_U$ for the local candidate for $(U,M_U)$.

\begin{lemma}[local bound]\label{lem:localbound}
For every nonempty set $U$, the local candidate $F_U$ for $(U,M_U)$ satisfies
\begin{equation}\label{eq:localbound}
 c(F_U)\le\alpha c(\cPs[U])+(\alpha+1)r_{\cPs}(U)d_U.
\end{equation}
\end{lemma}

\begin{proof}
As in Section~\ref{sec:overview}, link the $r_{\cPs}(U)$ fragments of $\cPs$ in $U$, in any cyclic order, into a closed walk by an edge from each exit to the next entry, if the two differ (Figure~\ref{fig:local}, left). Its vertices are exactly those of $U$, and it costs at most $c(\cPs[U])+r_{\cPs}(U)d_U$, so shortcutting it shows $c(C_U)\le\alpha\bigl(c(\cPs[U])+r_{\cPs}(U)d_U\bigr)$. Since $M_U$ has $m=r_{\cPs}(U)$ pairs, one for each fragment, \eqref{eq:candidatecost} then proves the bound.
\end{proof}

\subsection{The cost of the hybrid solution}

Let $\cU$ be a partition of $V$ into nonempty sets, and let $K$ be the multiset of edge uses of $\cPs$ between different parts. The hybrid solution $\Pt$ of $\cU$, described in Section~\ref{sec:interface}, uses the local candidates: in each part $U$, it replaces the fragments of $\cPs$ in $U$ by $F_U$, matching walks to fragments with the same pairs, and it retains the edge uses in $K$.

\begin{lemma}[hybrid bound]\label{lem:hybrid}
The hybrid solution is feasible and satisfies
\begin{equation}\label{eq:hybridbound}
 c(\Pt)\le\alpha\OPT-(\alpha-1)c(K)+(\alpha+1)\sum_{U\in\cU}r_{\cPs}(U)d_U.
\end{equation}
\end{lemma}

\begin{proof}
Each $F_U$ is a local solution for $(U,M_U)$, and replacing the fragments in one part does not change those in the other parts. So replacing them part by part keeps the family feasible, by Section~\ref{sec:local-solutions}. Also $\OPT=c(K)+\sum_Uc(\cPs[U])$. Applying Lemma~\ref{lem:localbound} gives
\begin{align*}
 c(\Pt)&=c(K)+\sum_Uc(F_U)\\
 &\le c(K)+\alpha\sum_Uc(\cPs[U])+(\alpha+1)\sum_U r_{\cPs}(U)d_U\\
 &=\alpha\OPT-(\alpha-1)c(K)+(\alpha+1)\sum_U r_{\cPs}(U)d_U.\qedhere
\end{align*}
\end{proof}

The retained edges are paid for at their actual cost, rather than multiplied by $\alpha$. With the \emph{allowance}
\begin{equation}\label{eq:allowance}
 D=\eps\OPT+(\alpha-1)c(K),
\end{equation}
the hybrid solution costs at most $(\alpha+\eps)\OPT$ whenever the partition satisfies the \emph{stopping condition}
\begin{equation}\label{eq:goal}
 (\alpha+1)\sum_{U\in\cU}r_{\cPs}(U)d_U\le D.
\end{equation}
Its left-hand side is the \emph{local error} of the partition. Once the stopping condition holds, the hybrid solution is cheap enough and no further refinement is needed; the analysis in Section~\ref{sec:process} refines the partition until this happens. Each edge use of $\cPs$ retained between parts increases the allowance $D$ by $\alpha-1$ times its cost; Section~\ref{sec:width} uses this increase to sharpen the bound on the number of fragments.

\section{The dynamic program}\label{sec:dp}

We now describe the algorithm. As in Section~\ref{sec:overview}, it is a dynamic program over vertex sets, and the sets are those that Section~\ref{sec:interface} makes enumerable: we write $\cR$ for the family of nonempty unions of at most $B_*$ intervals of $\sigma$, the order in which $\cQ$ lists the vertices, which by Lemma~\ref{lem:first-order} contains every set with at most $B_*$ fragments of $\cQ$. For each such set and each multiset of at most $B_*$ entry--exit pairs, it keeps the cheapest local solution it has found. The number $B_*$ is defined in~\eqref{eq:widthcap} in Section~\ref{sec:structure} and depends only on $\alpha$, $\eps$, the number $k$ of prescribed pairs, and the constants $\beta$ and $\ell$ of~\eqref{eq:qref}; it controls the size of the program (Section~\ref{sec:dp-time}). The program obtains local solutions in two ways: from one oracle tour of the set, as the local candidate of Section~\ref{sec:candidate}, and by combining local solutions of two disjoint smaller sets, guessing the edges between them. Its operations depend only on $c$, $M_0$, the oracle, $\cQ$, and $B_*$. Sections~\ref{sec:dp-states} and~\ref{sec:dp-combine} describe the program, Section~\ref{sec:dp-correct} compares its output with the hybrid solution of a partition, such as the one that the analysis constructs in Section~\ref{sec:process}, and Section~\ref{sec:dp-time} bounds its running time.

\subsection{States and local candidates}\label{sec:dp-states}

For each set $U\in\cR$, the program considers every nonempty multiset $M$ of at most $B_*$ ordered pairs of vertices of $U$, as a guess for the multiset $M_U$ of entry--exit pairs of the fragments of $\cPs$ in $U$; such a pair $(U,M)$ is a \emph{state}. The bound $B_*$ on the number of pairs suffices for the analysis, which uses the states $(U,M_U)$ only for sets $U$ with at most $B_*$ fragments of $\cPs$. The output comes from the state $(V,M_0)$, where $M_0$ is the multiset of prescribed pairs; for Theorem~\ref{thm:main}, this state is $(V,\{(s,t)\})$. For every state $(U,M)$, the table cell $Z[U,M]$ keeps the cheapest local solution for $(U,M)$, as defined in Section~\ref{sec:local-solutions}, that the program has found so far. To \emph{record} a family for $(U,M)$ means to keep it in $Z[U,M]$ if the cell is still empty or holds a family of larger total cost. Sets are processed in increasing order of cardinality, so the families kept for smaller sets are final when they are combined.

For every set $U\in\cR$, the program calls the oracle once to obtain the oracle tour $C_U$ of Section~\ref{sec:candidate} and records the local candidate for every state $(U,M)$; the same tour $C_U$ is used for all $M$. So every cell holds a local solution as soon as its set has been processed.

\subsection{Combining two sets}\label{sec:dp-combine}

Let $(U,M)$ be a state with $M=\{(a_1,b_1),\ldots,(a_m,b_m)\}$, and let $U=U_1\mathbin{\dot\cup}U_2$ with $U_1,U_2\in\cR$. The program goes through all states $(U_1,M_1)$ and $(U_2,M_2)$ whose table cells are nonempty, and all multisets $L$ of at most $2B_*$ edges between $U_1$ and $U_2$, in either direction; the same edge may occur several times in $L$. As in Section~\ref{sec:overview}, in each case it joins the walks kept for $(U_1,M_1)$ and $(U_2,M_2)$ by the edges of $L$: it tries every way to concatenate these walks and the edges of $L$ into walks, using each kept walk exactly once and each edge as often as it occurs in $L$. If the resulting walks are $m$ walks, one from $a_i$ to $b_i$ for each $i$, the family is recorded for $(U,M)$; otherwise it is discarded. Every kept walk is used because it may be the only one covering some vertex. Figure~\ref{fig:combine} shows an example.

Keeping only a cheapest family for each state loses nothing. By Section~\ref{sec:local-solutions}, in any of these concatenations the family kept for $(U_1,M_1)$ can be replaced by any other local solution for $(U_1,M_1)$, and likewise for $(U_2,M_2)$, without changing the end vertices of the resulting walks; a cheaper family for $(U_1,M_1)$ or $(U_2,M_2)$ therefore gives a cheaper family for $(U,M)$.

\medskip
\begin{algorithm}[H]\small
\caption{\textup{\textsc{PrescribedPaths}}$(c,M_0,\mathcal A,\cQ,B_*)$}
\label{alg:main}
\KwIn{a directed metric $c$ on $V$, a multiset $M_0$ of $k$ prescribed pairs, an oracle $\mathcal A$, a family $\cQ$ of vertex-disjoint paths that together visit every vertex, and the bound $B_*\ge k$.}
\KwOut{a feasible family.}
\BlankLine
Form the order $\sigma$ of $\cQ$ and the family $\cR$ of nonempty unions of at most $B_*$ intervals of $\sigma$\;
Initialize all cells of $Z$ as empty\;
\For{each $U\in\cR$, in increasing order of cardinality}{
  Call $\mathcal A$ once on $U$, obtaining $C_U$\;
  \For{each state $(U,M)$}{
    Record the local candidate for $(U,M)$ built from $C_U$\;
    \For{each partition $U=U_1\mathbin{\dot\cup}U_2$ with $U_1,U_2\in\cR$}{
      \For{each pair of states $(U_1,M_1)$, $(U_2,M_2)$ with nonempty table cells and each multiset $L$ of at most $2B_*$ edges between $U_1$ and $U_2$}{
        For each concatenation of the walks kept for $(U_1,M_1)$ and $(U_2,M_2)$ and the edges of $L$ into walks, using each kept walk once and each edge as often as it occurs in $L$, record the resulting family for $(U,M)$ if its walks have exactly the pairs of $M$\;
      }
    }
  }
}
Return the family kept in $Z[V,M_0]$\;
\end{algorithm}
\medskip

\subsection{Comparison with hybrid solutions}\label{sec:dp-correct}

\begin{lemma}[soundness]\label{lem:dp-sound}
Every family that the program records for a state $(U,M)$ is a local solution for $(U,M)$. In particular, the output, the family kept in $Z[V,M_0]$, is a feasible family of the instance.
\end{lemma}

\begin{proof}
We use induction over increasing set size. The local candidates are local solutions (Section~\ref{sec:candidate}). A family recorded by combining two sets has exactly the pairs of $M$, since all other concatenations are discarded. Its walks lie in $U$, and they cover $U$, since they contain all walks kept for $(U_1,M_1)$ and $(U_2,M_2)$, which by induction cover $U_1$ and $U_2$. Finally, $V$ is a single interval of $\sigma$, so $V\in\cR$, and $M_0$ has $k\le B_*$ pairs, so $(V,M_0)$ is a state. Since the local candidate is recorded for every state, the cell $Z[V,M_0]$ holds a local solution for $(V,M_0)$, which is a feasible family.
\end{proof}

As in Section~\ref{sec:overview}, the analysis compares the output with the hybrid solution of a partition obtained from $V$ by repeated divisions of a set into two. It records these divisions in a \emph{refinement tree}: a rooted tree whose nodes are nonempty vertex sets, such that the root is $V$ and every node either is a leaf, without children, or has two children that partition it. The leaves of a refinement tree form a partition of $V$; if $V$ is not divided, the root is the only leaf. Suppose that every node of the tree has at most $B_*$ fragments of each of $\cPs$ and $\cQ$. Then the dynamic program can follow the tree from the leaves to the root, for two reasons. Since every node has few fragments of $\cQ$, it is a union of few intervals of $\sigma$ (Lemma~\ref{lem:first-order}), so it is one of the sets in $\cR$ that the program considers. Since every node has few fragments of $\cPs$, the entry--exit pairs of $\cPs$ in it form a state, so the program considers the pairs that the hybrid solution uses inside the node. The following theorem shows that the program then finds a solution no more expensive than the hybrid solution of the leaves, although it does not know the tree.

\begin{theorem}[the dynamic program follows refinement trees]\label{thm:dp}
Let $\cH$ be a refinement tree with set of leaves $\cU$. If every node $W$ of $\cH$ has at most $B_*$ fragments of each of $\cPs$ and $\cQ$, then Algorithm~\ref{alg:main} returns a feasible family whose cost is at most that of the hybrid solution of $\cU$.
\end{theorem}

\begin{proof}
As explained before the theorem, every node $W$ of $\cH$ belongs to $\cR$, and $(W,M_W)$ is a state, where $M_W$ is the multiset of entry--exit pairs of the fragments of $\cPs$ in $W$. Let $\Pt$ be the hybrid solution of $\cU$, and for a node $W$ let $\Pt_W$ be the family of fragments of $\Pt$ in $W$. The edges of $\Pt$ are the edge uses of $\cPs$ between different leaves and the edges of walks inside the leaves. Since $W$ is a union of leaves, $\Pt$ enters and leaves $W$ along the same edge uses as $\cPs$, so the walks of $\Pt_W$ have exactly the pairs of $M_W$. Writing $c(Z[W,M_W])$ for the cost of the family that the program finally keeps in $Z[W,M_W]$, we show, from the leaves up to the root, that every node $W$ satisfies
\begin{equation}\label{eq:dp-claim}
 c(Z[W,M_W])\le c(\Pt_W).
\end{equation}

At a leaf $U$, the family $\Pt_U$ is the local candidate $F_U$ for $(U,M_U)$, which the program records, so~\eqref{eq:dp-claim} holds.

Now let $W$ have children $W_1,W_2$, and suppose that~\eqref{eq:dp-claim} holds for $W_1$ and $W_2$. Let $L$ be the multiset of edge uses of $\cPs$ between $W_1$ and $W_2$. These edges join different leaves, so they are exactly the edges of $\Pt_W$ between $W_1$ and $W_2$. Cutting them splits the walks of $\Pt_W$ into those of $\Pt_{W_1}$ and $\Pt_{W_2}$, and each cut adds one walk. Since $W_1$ and $W_2$ have at most $B_*$ fragments of $\cPs$ each, this gives
\[
 |L|=r_{\cPs}(W_1)+r_{\cPs}(W_2)-r_{\cPs}(W)\le2B_*.
\]
Thus $\Pt_W$ is a concatenation of the walks of $\Pt_{W_1}$ and $\Pt_{W_2}$ and the edges of $L$, using each walk once and each edge as often as it occurs in $L$. Since $|L|\le2B_*$, the program enumerates $L$ for the partition $W=W_1\mathbin{\dot\cup}W_2$ and the states $(W_1,M_{W_1})$ and $(W_2,M_{W_2})$. In the concatenation, replace the walks of $\Pt_{W_1}$ and $\Pt_{W_2}$ by the walks kept for these two states, which are final when $W$ is processed, matching walks with the same pairs. The concatenation stays valid, its walks keep the pairs of $M_W$, and by~\eqref{eq:dp-claim} for $W_1$ and $W_2$ its cost is at most
\[
 c(Z[W_1,M_{W_1}])+c(Z[W_2,M_{W_2}])+c(L)\le c(\Pt_{W_1})+c(\Pt_{W_2})+c(L)=c(\Pt_W).
\]
The program records it for $(W,M_W)$, so~\eqref{eq:dp-claim} holds for $W$.

At the root, $\Pt_V=\Pt$ and $M_V=M_0$, so the family kept in $Z[V,M_0]$ costs at most $c(\Pt)$. By Lemma~\ref{lem:dp-sound}, it is a feasible family.
\end{proof}

So the analysis only needs a suitable refinement tree: if the leaves of a tree as in Theorem~\ref{thm:dp} also satisfy the stopping condition~\eqref{eq:goal}, then their hybrid solution costs at most $(\alpha+\eps)\OPT$ by Lemma~\ref{lem:hybrid}, and so does the output. Section~\ref{sec:structure} states that such a tree exists.

\subsection{Running time}\label{sec:dp-time}

\begin{lemma}[running time]\label{lem:dp-time}
The dynamic program calls the oracle at most $(n+1)^{2B_*}$ times and, apart from these calls, runs in time $(n+1)^{O(B_*\log(B_*+1))}$ times a polynomial in the input size.
\end{lemma}

\begin{proof}
Each set in $\cR$ is determined by the at most $2B_*$ endpoints of its intervals, so $\cR$ has at most $(n+1)^{2B_*}$ sets, and the program calls the oracle once for each of them. Each set has at most $(n^2+1)^{B_*}$ states. For each state, the program chooses $U_1\in\cR$, which determines $U_2$, two states, and a multiset $L$, in at most $(n+1)^{2B_*}(n^2+1)^{4B_*}$ ways. In each case, the at most $4B_*$ kept walks and edges of $L$ can be concatenated in at most $(4B_*)!\,2^{4B_*}$ ways. Each concatenation is formed and compared in time polynomial in the input size. Since $n\ge2$, the factor $(4B_*)!\,2^{4B_*}\le(8B_*)^{4B_*}$ is at most $(n+1)^{O(B_*\log(B_*+1))}$, and multiplying these numbers gives the bound.
\end{proof}

Since $B_*$ is a constant, the dynamic program makes polynomially many oracle calls and performs polynomial additional work, so it runs in polynomial time when the oracle does. The reference family is also computed in polynomial time: for Asymmetric Path TSP by one run of the Feige--Singh reduction (Section~\ref{sec:interface}), and in general as in Appendix~\ref{sec:initialization}.

\section{The structural theorem and the proofs of Theorems~\ref{thm:main} and~\ref{thm:unified}}\label{sec:structure}

It remains to find a refinement tree as in Theorem~\ref{thm:dp} whose leaves satisfy the stopping condition~\eqref{eq:goal}. In this section we define $B_*$, state that such a tree exists, and derive Theorems~\ref{thm:main} and~\ref{thm:unified}. The existence statement is proved in Sections~\ref{sec:fragments} to~\ref{sec:width}.

As in Section~\ref{sec:overview}, we count the fragments of $\cPs$ and $\cQ$ together, but the proof gives the fragments of $\cPs$ a larger weight. Cutting an edge of $\cPs$ also increases the allowance $D$ of~\eqref{eq:allowance}, whereas cutting an edge of $\cQ$ does not, so the proof lets the cut edges of $\cQ$ count less. For a positive integer $A$ defined below, the \emph{width} of a nonempty set $U$ is
\begin{equation}\label{eq:width}
 w(U)=Ar_{\cPs}(U)+r_{\cQ}(U).
\end{equation}
Since $A\ge1$, both $r_{\cPs}(U)$ and $r_{\cQ}(U)$ are at most $w(U)$; a bound on these two numbers is all that the dynamic program needs. Their maximum would serve the dynamic program as well; we use the weighted sum because it is convenient in the analysis. The fragments of $\cPs$ and $\cQ$ in $V$ are their walks, $k$ and at most $\ell$ of them, so $w(V)\le Ak+\ell$.

Define the integers
\begin{equation}\label{eq:A}
 A=\left\lceil\frac{8(\alpha+1)(1+\beta)}{\eps}\right\rceil,
 \qquad
 B_0=Ak+\ell,
\end{equation}
\begin{equation}\label{eq:widthcap}
 p=\left\lceil\frac{8(\alpha+1)}{\alpha-1}\right\rceil,
 \qquad
 B_*=(2AB_0)^{\lceil3(1+(\alpha-1)/\eps)^p\rceil}.
\end{equation}
They depend only on $\alpha,\beta,\eps,k,\ell$. The integer $B_0$ bounds the initial width, and $B_*$ is the bound of the following theorem. The values of $A$, $p$ and $B_*$ come from its proof in Sections~\ref{sec:process} and~\ref{sec:width}, where they are explained.

\begin{theorem}[the structural theorem]\label{thm:process}
There is a refinement tree in which every node $W$ has width $w(W)<B_*$ and whose leaves satisfy the stopping condition~\eqref{eq:goal}.
\end{theorem}

For fixed $\alpha$, the exponent $p$ is a constant, and for $0<\eps\le1$, \eqref{eq:A} and~\eqref{eq:widthcap} give
\begin{equation}\label{eq:singleexponential}
 \ln B_*=
 O_\alpha\left(
 \eps^{-p}\log\frac{(\beta+1)(k+\ell+1)}{\eps}
 \right).
\end{equation}
Thus $B_*$ is singly exponential in a polynomial in $1/\eps$; the exponent $p$ increases as $\alpha$ approaches $1$.

\subsection{Proofs of Theorems~\ref{thm:main} and~\ref{thm:unified}}\label{sec:proof-interface}

\begin{lemma}[any oracle and reference family]\label{lem:general}
For every oracle as in Section~\ref{sec:candidate} and every reference family satisfying~\eqref{eq:qref}, Algorithm~\ref{alg:main}, run with $B_*$ as in~\eqref{eq:widthcap}, returns a feasible family of cost at most $(\alpha+\eps)\OPT$. It makes polynomially many oracle calls and performs polynomial additional work; for fixed $\alpha$ and $0<\eps\le1$, the degree of the polynomial is
\begin{equation}\label{eq:interface-time}
 \exp\left(O_\alpha\left(\eps^{-p}\log\frac{(\beta+1)(k+\ell+1)}{\eps}\right)\right).
\end{equation}
\end{lemma}

\begin{proof}
We have $B_*>B_0\ge k$, as required by Algorithm~\ref{alg:main}. By Theorem~\ref{thm:process}, some refinement tree has nodes of width less than $B_*$, and hence with fewer than $B_*$ fragments of each of $\cPs$ and $\cQ$, and leaves that satisfy the stopping condition~\eqref{eq:goal}. By Theorem~\ref{thm:dp} and Lemma~\ref{lem:hybrid}, the output is feasible and costs at most $(\alpha+\eps)\OPT$. The running time follows from Lemma~\ref{lem:dp-time} and~\eqref{eq:singleexponential}.
\end{proof}

\begin{proof}[Proof of Theorem~\ref{thm:main}]
For Asymmetric Path TSP, we have $k=1$, and the path of the Feige--Singh reduction serves as $\cQ$, with $\beta=3\alpha$ and $\ell=1$, as in Section~\ref{sec:overview}. Lemma~\ref{lem:general} gives an $s$-$t$ walk that visits every vertex and costs at most $(\alpha+\eps)\OPT$. Shortcutting it gives the claimed approximation. Alternatively, Lemma~\ref{lem:capped} with $k=1$ provides a reference family with $\beta=3\alpha$ and $\ell=\lceil2\alpha\rceil$, so the theorem also follows from Lemma~\ref{lem:general} without the Feige--Singh reduction.
\end{proof}

\begin{proof}[Proof of Theorem~\ref{thm:unified}]
Lemma~\ref{lem:capped} constructs a reference family with $\beta=\alpha(2k+1)$ and $\ell=\lceil\alpha(k+1)\rceil$ in polynomial time, and Lemma~\ref{lem:general} gives the claimed approximation. Substituting these bounds into~\eqref{eq:interface-time} shows that, for fixed $\alpha$ and $0<\eps\le1$, the degree is $\exp(O_\alpha(\eps^{-p}\log((k+1)/\eps)))$.
\end{proof}

\section{Dividing along levels}\label{sec:fragments}

Sections~\ref{sec:fragments} to~\ref{sec:width} prove Theorem~\ref{thm:process} by constructing the refinement tree. The tree starts with $V$ alone and grows step by step until the stopping condition holds: dividing a leaf attaches two children to it, which become leaves, while the divided set stays in the tree as an inner node. So the leaves change during the construction, and they are always the current parts; Theorem~\ref{thm:process} concerns the leaves of the final tree. This section describes how a part is divided at one or several levels, how the division is recorded in the tree, and how it changes the width. Section~\ref{sec:process} then chooses the levels, and Section~\ref{sec:width} adds up the payments for the growth of the width.

As in Section~\ref{sec:overview}, let $U$ be a set of positive diameter $d_U$ (Section~\ref{sec:candidate}), choose an ordered pair $(a_U,b_U)$ with $c(a_U,b_U)=d_U$, and give every vertex $v\in U$ the height $f_U(v)=c(a_U,v)\in[0,d_U]$. A level $\theta\in[0,d_U)$ divides $U$ into the \emph{lower set} $\{v\in U:f_U(v)\le\theta\}$, which contains $a_U$, and the \emph{upper set} of the vertices of height above $\theta$, which contains $b_U$ (Figure~\ref{fig:division}). When a leaf $U$ of the tree is divided at $\theta$, these two sets become the children of $U$, its lower and upper child. Dividing at several levels creates slabs: a \emph{slab} of $U$ consists of the vertices of $U$ whose heights lie in $(\theta,\theta']$ for two levels $\theta<\theta'$, or are at most $\theta$, or are above $\theta$, for one level $\theta$; these one or two levels are its \emph{boundary levels}. An edge use $(u,v)$ inside $U$ \emph{crosses $\theta$ upward} if $f_U(u)\le\theta<f_U(v)$. For a multiset $F$ of edges inside $U$, the \emph{upward crossing number} $\up_F(\theta)$ is the number of edge uses in $F$ that cross $\theta$ upward, counting multiplicities; for a family of walks, we count all of their edge uses.

As in Figure~\ref{fig:division}, the upward crossing numbers at its boundary levels bound how many fragments a slab can gain.

\begin{lemma}[slab bound]\label{lem:fragments}
Let $\cP$ be a family of walks in $U$. For every slab $Y$ of $U$,
\[
 r_{\cP}(Y)\le|\cP|+\sum_\theta\up_{\cP}(\theta),
\]
where the sum is over the boundary levels $\theta$ of $Y$.
\end{lemma}

\begin{proof}
Between two consecutive fragments in $Y$, a walk makes an excursion outside $Y$. The excursion reaches a vertex above an upper boundary level of $Y$, and then crosses this level upward on the way out, or a vertex at or below a lower boundary level, and then crosses this level upward on the way back. Hence every walk has at most one more fragment in $Y$ than it has upward crossings of the boundary levels of $Y$, and summing over the walks of $\cP$ proves the claim.
\end{proof}

Lemma~\ref{lem:fragments} also bounds the width~\eqref{eq:width} of a slab $Y$ of $U$. Apply it to the family of $A$ copies of each fragment of $\cPs$ in $U$ and one copy of each fragment of $\cQ$ in $U$, which has $w(U)$ walks. Since $Y\subseteq U$, the fragments of this family in $Y$ are the fragments of $\cPs$ and $\cQ$ in $Y$, with the same multiplicities, so there are $w(Y)$ of them. With the \emph{weighted upward crossing number}
\begin{equation}\label{eq:weightedup}
 u_U(\theta)=A\up_{\cPs[U]}(\theta)+\up_{\cQ[U]}(\theta),
\end{equation}
where $\cPs[U]$ and $\cQ[U]$ are the multisets of edge uses of $\cPs$ and $\cQ$ inside $U$ (Section~\ref{sec:interface}), every slab $Y$ of $U$ therefore satisfies
\begin{equation}\label{eq:slabwidth}
 w(Y)\le w(U)+\sum_\theta u_U(\theta),
\end{equation}
where the sum is over the boundary levels $\theta$ of $Y$.

The analysis may divide a leaf $U$ of the tree at several levels $\theta_1<\cdots<\theta_m$ at once, chosen so that all the sets created are nonempty. The tree records this by binary divisions, as in Figure~\ref{fig:division}: first $U$ is divided at $\theta_1$, then its upper child at $\theta_2$, and so on, so that the new leaves are the vertices of height at most $\theta_1$, those with heights in $(\theta_{i-1},\theta_i]$ for $2\le i\le m$, and those of height above $\theta_m$. Every set added to the tree is a slab of $U$ whose one or two boundary levels are among $\theta_1,\ldots,\theta_m$. So, by~\eqref{eq:slabwidth}, its width exceeds $w(U)$ by at most the weighted upward crossing numbers of two chosen levels, however many levels are chosen.

The upward crossing numbers thus control the growth of the width. They are in turn controlled by cost: integrated over all levels, they are at most the cost of the edges (Figure~\ref{fig:coarea}).

\begin{lemma}[cost bounds upward crossings]\label{lem:coarea}
For every multiset $F$ of edges inside $U$,
\[
 \int_0^{d_U}\up_F(\theta)\,d\theta\le c(F).
\]
\end{lemma}

\begin{proof}
An edge use $(u,v)$ with $f_U(u)<f_U(v)$ crosses exactly the levels in $[f_U(u),f_U(v))$ upward, an interval of length $f_U(v)-f_U(u)\le c(u,v)$ by the triangle inequality $c(a_U,v)\le c(a_U,u)+c(u,v)$. Other edge uses cross no level upward. Summing the lengths of these intervals over the edge uses of $F$ proves the inequality.
\end{proof}

As explained in Section~\ref{sec:overview}, this is what allows the analysis to divide at many levels at once: the growth of the width of each set created depends only on its at most two boundary levels, whereas, by Lemma~\ref{lem:coarea}, the edges that cross some level of a set $\Theta$ of levels upward cost at least the integral of the upward crossing numbers over all of $\Theta$.

\section{Choosing the levels}\label{sec:process}

In this section we choose the levels at which the parts are divided; this completes the process that constructs the tree of Theorem~\ref{thm:process}. The process depends on $\cPs$ and is used only in the analysis. Along with the tree, it maintains a bound $B$ on the width of the sets of the tree; $B$ starts at $B_0$ and may increase from step to step, but never decreases. Each step processes parts as in Section~\ref{sec:overview}, in one of two ways. A \emph{free step} divides one part by a free split, at a single level without upward crossings, and leaves $B$ unchanged. A \emph{paid step} divides the parts at many levels at once and increases $B$; the main result of this section, Lemma~\ref{lem:refinement-step}, shows that the edges it cuts pay for this increase. Section~\ref{sec:width} then shows that $B$ stays below the constant $B_*$ of~\eqref{eq:widthcap}; since $B$ never decreases, it suffices to bound its final value.

\subsection{The process and its invariant}

The process maintains a refinement tree $\cH$, as defined in Section~\ref{sec:dp-correct}, whose leaves form the current partition $\cU$ of $V$, together with the integer $B$. The invariant is
\begin{equation}\label{eq:treewidth}
 w(W)\le B\qquad\text{for every set $W$ in $\cH$}.
\end{equation}
Since $B$ never decreases and the sets already in $\cH$ do not change, a step preserves~\eqref{eq:treewidth} as soon as the sets it adds satisfy it. Initially $\cU=\{V\}$, $\cH$ has just its root, and $B=B_0$, which bounds $w(V)$ by Section~\ref{sec:structure}. Let $K$ be the multiset of edge uses of $\cPs$ between different parts and set $D=\eps\OPT+(\alpha-1)c(K)$, as in~\eqref{eq:allowance}. This is the allowance of the stopping condition~\eqref{eq:goal}: the process may stop once the local error is at most $D$, so a larger $D$ brings it closer to its goal.

In each step, proceed as follows. If the stopping condition~\eqref{eq:goal} holds, stop. If it fails, perform a free step or a paid step, as described next.

For each step, write $B$ and $D$ for their values immediately before the step and $B'$ and $D'$ for their values immediately after it. Let $E_P$ and $E_Q$ be the multisets of newly cut edge uses of $\cPs$ and $\cQ$: their endpoints are together before the step and in different parts afterwards. The edge uses in $E_P$ join $K$, so~\eqref{eq:allowance} gives
\begin{equation}\label{eq:Dupdate}
 D'=D+(\alpha-1)c(E_P).
\end{equation}
A free step may also increase $D$.

\subsection{Free steps and paid steps}\label{sec:refinement-step}

From now on, $U$ denotes a part of the current partition $\cU$, a leaf of the tree, and sums over $U$ run over all parts. The \emph{measure} $\meas\Theta$ of a set $\Theta$ of levels of $U$ that is a finite union of intervals is its total length (Section~\ref{sec:overview}); we measure the sets of levels of each part on its own height axis, and add these measures when considering several parts. If the stopping condition fails, the parts have a large total diameter, so the levels along which they can be divided have large total measure. Indeed, if $(\alpha+1)\sum_U r_{\cPs}(U)d_U>D$, then, since $Ar_{\cPs}(U)\le w(U)\le B$ by~\eqref{eq:treewidth},
\begin{equation}\label{eq:diamlength}
 \sum_{U\in\cU}d_U>
 \frac{AD}{(\alpha+1)B}.
\end{equation}
Every part $U$ with $d_U>0$ has the heights $f_U$ and the levels $\theta\in[0,d_U)$ of Section~\ref{sec:fragments}, and the weighted upward crossing numbers $u_U$ of~\eqref{eq:weightedup}; a part of diameter zero has no levels.

\paragraph{Free step.}
If some level $\theta$ of some part $U$ has $u_U(\theta)=0$, divide $U$ at $\theta$ alone: the lower and upper sets of $\theta$ become the two children of $U$. This is a free split as in Section~\ref{sec:overview}, and $B$ stays unchanged. Both sets are nonempty, since the lower set contains $a_U$ and the upper set contains $b_U$, and by~\eqref{eq:slabwidth} neither has larger width than $U$, so~\eqref{eq:treewidth} is preserved.

\paragraph{Paid step.}
Otherwise every level of every part has at least one weighted upward crossing, and we allow $B$ to increase. Then $D>0$. Indeed, since the stopping condition fails, some part has positive diameter; its levels are crossed upward by edge uses of $\cPs$ or $\cQ$, which have positive cost by the triangle inequality; and $c(\cQ)\le\beta\OPT$. So $\OPT>0$, and $D\ge\eps\OPT>0$ by~\eqref{eq:allowance}. Suppose that, for some $h\ge1$, we divide the parts only at levels whose weighted upward crossing numbers lie in $[h,2h)$. This has two effects, both proportional to $h$.

First, $B$ has to increase only by $4h$. As shown at the end of Section~\ref{sec:fragments}, every set created from a part $U$ is a slab of $U$ whose at most two boundary levels are among the levels used. Each of them has fewer than $2h$ weighted upward crossings, so~\eqref{eq:slabwidth}, the form of Lemma~\ref{lem:fragments} for the width, and $w(U)\le B$ give the set width less than $B+4h$, however many levels are used.

Second, the cut edges cost at least $h$ per unit of measure. The edges crossing the levels used upward are cut, and by Lemma~\ref{lem:coarea} their weighted cost, in which edge uses of $\cPs$ count $A$ times, is at least $h$ times the measure of the set of levels used.

Comparing the two effects, the weighted cost cut per unit increase of $B$ is at least a quarter of this measure, whatever $h$ is. For a larger $h$, $B$ increases more, but each level used is crossed upward more often, and the cut edges cost proportionally more; so the values of $h$ differ only in the measure of the levels they allow. Since $B$ bounds the width of the sets of all parts, the step uses the same $h$ in all parts, and the cut edges of all parts together pay for the single increase of $B$. What has to be large is therefore the total measure, over all parts, of the levels used. We find such an $h$, and divide at all of these levels, in three stages.

\emph{Discarding levels with many crossings.} The total amount of crossings is bounded: Lemma~\ref{lem:coarea}, applied in each part, gives
\begin{equation}\label{eq:weightedcoarea}
 \sum_U\int_0^{d_U}u_U(\theta)\,d\theta
 \le A c(\cPs)+c(\cQ)\le(A+\beta)\OPT.
\end{equation}
We discard levels with $u_U(\theta)>AB$, so that the remaining weighted upward crossing numbers fall into logarithmically many ranges $[h,2h)$. By~\eqref{eq:weightedcoarea}, the discarded levels have total measure at most $(A+\beta)\OPT/(AB)\le(1+\beta)\OPT/B$. Since $A\ge4(\alpha+1)(1+\beta)/\eps$ by~\eqref{eq:A} (the factor $8$ there, rather than $4$, is used only in Section~\ref{sec:width}) and $D\ge\eps\OPT$, this is at most $AD/(4(\alpha+1)B)$, a quarter of the lower bound~\eqref{eq:diamlength}. Hence the remaining levels still have total measure at least $3AD/(4(\alpha+1)B)$.

\emph{Choosing one range.} Every remaining level has a weighted upward crossing number between $1$ and $AB$, so, as in Section~\ref{sec:overview}, it lies in exactly one of the $b(B)=1+\lfloor\log_2(AB)\rfloor$ ranges $[h,2h)$ with $h=2^0,\ldots,2^{b(B)-1}$. For such an $h$ and a part $U$, let
\[
 \Theta_U(h)=\{\theta\in[0,d_U):h\le u_U(\theta)<2h,\ u_U(\theta)\le AB\}
\]
be the remaining levels of $U$ in the range for $h$. The step uses, for all parts, the value of $h$ with the largest total $\sum_U\meas\Theta_U(h)$. Since the largest total is at least the average of the $b(B)$ totals,
\begin{equation}\label{eq:bucketlength}
\begin{split}
 \sum_U\meas\Theta_U(h)
 &\ge\frac1{b(B)}\sum_{j=0}^{b(B)-1}\sum_U\meas\Theta_U(2^j)
 =\frac1{b(B)}\sum_U\meas\{\theta\in[0,d_U):u_U(\theta)\le AB\}\\
 &\ge\frac{3AD}{4(\alpha+1)B\,b(B)},
\end{split}
\end{equation}
where the equality holds because every remaining level lies in exactly one range, and the last inequality is the bound at the end of the previous stage.
Only this total is guaranteed to be large, and only the total enters the payment; in a single part, $\Theta_U(h)$ may be small or even empty. The number $b(B)$ of ranges is the only loss in this stage.

\emph{Dividing at all levels of this range.} Let $\eta<\eta'$ be two consecutive vertex heights of $U$. All levels $\theta\in[\eta,\eta')$ have the same lower set, the vertices of height at most $\eta$, and hence give the same division of $U$; they also have the same weighted upward crossing number. So each set $\Theta_U(h)$ is a union of such intervals $[\eta,\eta')$, and, as in Section~\ref{sec:overview} (Figure~\ref{fig:coarea}), it suffices to divide $U$ at one arbitrary level of each of them. An edge use that crosses some level of $[\eta,\eta')$ upward runs from a vertex of height at most $\eta$ to one of height at least $\eta'$, so it also crosses the chosen level upward and is cut. Chosen levels in different intervals have a vertex height between them, so all sets created are nonempty. Divide every part $U$ at all chosen levels and record the division by binary divisions, as in Figure~\ref{fig:division}. The bound becomes
\begin{equation}\label{eq:Bupdate}
 B'=B+4h.
\end{equation}

\begin{lemma}[progress and payment]\label{lem:refinement-step}
Every step divides at least one part and preserves the invariant~\eqref{eq:treewidth}. In a paid step,
\begin{equation}\label{eq:cutpayment}
 c(E_P)+\frac{c(E_Q)}A
 \ge
 \frac{3D}{4(\alpha+1)}\frac{h}{B\,b(B)}.
\end{equation}
\end{lemma}

\begin{proof}
For free steps, both claims were shown above. In a paid step, \eqref{eq:bucketlength} shows that $\Theta_U(h)$ is nonempty for some part $U$, so at least one part is divided. All chosen levels have weighted upward crossing numbers in $[h,2h)$, so, as shown at the beginning of the paid step, every new set of the tree has width less than $B+4h=B'$, and~\eqref{eq:treewidth} is preserved.

For the payment, fix a part $U$ and let $E_U$ be the multiset of newly cut edge uses inside $U$, in which those of $\cPs$ are taken $A$ times. For $\theta\in\Theta_U(h)$, every edge use of $\cPs$ or $\cQ$ inside $U$ that crosses $\theta$ upward is newly cut, as shown when the levels were chosen, so $\up_{E_U}(\theta)=u_U(\theta)$ by~\eqref{eq:weightedup}. Lemma~\ref{lem:coarea}, applied to $E_U$, gives
\[
 c(E_U)\ge\int_0^{d_U}\up_{E_U}(\theta)\,d\theta\ge\int_{\Theta_U(h)}u_U(\theta)\,d\theta\ge h\meas\Theta_U(h).
\]
Each newly cut edge use lies inside one part before the step, so summing over all parts and using~\eqref{eq:bucketlength} gives
\[
 A c(E_P)+c(E_Q)=\sum_U c(E_U)\ge h\sum_U\meas\Theta_U(h)\ge\frac{3AD}{4(\alpha+1)}\frac{h}{B\,b(B)}.
\]
Divide by $A$ to obtain~\eqref{eq:cutpayment}.
\end{proof}

Lemma~\ref{lem:refinement-step} says that every increase of $B$ is paid for. By~\eqref{eq:Bupdate}, a paid step increases $B$ by the relative amount $4h/B$, and~\eqref{eq:cutpayment} bounds it by the newly cut cost relative to the allowance:
\begin{equation}\label{eq:relgrowth}
 \frac{4h}{B}\le\frac{16(\alpha+1)}{3}\,b(B)\,\frac{c(E_P)+c(E_Q)/A}{D}.
\end{equation}
Thus $h/B$ is controlled: $B$ grows by a large factor only if the step cuts edges of $\cPs$ and $\cQ$ whose cost is a large fraction of $D$, up to the factor $b(B)$ that comes from the number of ranges.

\section{Adding up the payments}\label{sec:width}

In this section we bound $B$ over the whole process and complete the proof of Theorem~\ref{thm:process}. If $\OPT=0$, then every step is a free step by Section~\ref{sec:refinement-step}, so $B=B_0<B_*$ throughout. From now on we assume $\OPT>0$. Each edge use of $\cPs$ cut by the process has two effects: it pays for an increase of $B$, and it increases the allowance $D$. Section~\ref{sec:firstbound} uses only the first effect, which already bounds the width independently of the input size; Section~\ref{sec:potential} uses both and obtains the bound $B_*$. The weight $A$ in the width~\eqref{eq:width} makes the total effect of the cut edges of $\cQ$ small enough that it changes the bound on $\log B$ only by a constant factor.

Recall from Section~\ref{sec:process} that $E_P$ and $E_Q$ are the multisets of edge uses of $\cPs$ and $\cQ$ newly cut in a step, and that a paid step pays $c(E_P)+c(E_Q)/A$ for the increase of $B$.

\begin{lemma}[each edge use is cut at most once]\label{lem:fresh}
Summed over all steps of the process,
\[
 \sum c(E_P)\le\OPT,\qquad \sum c(E_Q)\le\beta\OPT.
\]
\end{lemma}

\begin{proof}
Parts are only divided, never merged. Once the endpoints of an edge use lie in different parts, they remain in different parts. The cost bounds follow by summing over the original edge uses of $\cPs$ and $\cQ$.
\end{proof}

\subsection{A first bound}\label{sec:firstbound}

The factor $b(B)$ in~\eqref{eq:relgrowth} is logarithmic in $B$, so it is the logarithm of $B$ whose relative growth is bounded by the payment divided by $D$, up to a constant. We therefore track $B$ through
\[
 L(B)=\ln(2AB),
\]
which is chosen so that the number of ranges satisfies $b(B)=1+\lfloor\log_2(AB)\rfloor\le\log_2(2AB)=L(B)/\ln2$.

\begin{lemma}[growth of the logarithm]\label{lem:logstep}
In every paid step,
\begin{equation}\label{eq:singlelogstep}
 \frac{L(B')-L(B)}{L(B)}
 \le\frac{8(\alpha+1)}{D}
       \left(c(E_P)+\frac{c(E_Q)}A\right).
\end{equation}
\end{lemma}

\begin{proof}
Since $B$ grows by $4h$, $\ln B$ grows by at most $4h/B$. Since $b(B)\le L(B)/\ln2$, \eqref{eq:relgrowth} gives
\[
 \frac{L(B')-L(B)}{L(B)}
 =\frac{\ln(1+4h/B)}{L(B)}
 \le\frac{4h}{B L(B)}
 \le\frac{16(\alpha+1)}{3\ln2}\,
       \frac{c(E_P)+c(E_Q)/A}{D}.
\]
The claim follows from $\ln2>2/3$.
\end{proof}

Since $D\ge\eps\OPT$, multiplying~\eqref{eq:singlelogstep} over all paid steps and using $1+x\le\ee^x$ and Lemma~\ref{lem:fresh} gives
\[
 \ln(2AB)
 \le\ln(2AB_0)
 \exp\left(\frac{8(\alpha+1)}{\eps}\Bigl(1+\frac{\beta}{A}\Bigr)\right)
 \le\ln(2AB_0)\,\ee^{(\alpha+1)(8/\eps+1)},
\]
where the last step uses $\beta/A\le\eps/8$, by~\eqref{eq:A}. Exponentiating gives
\[
 B<2AB\le(2AB_0)^{\ee^{(\alpha+1)(8/\eps+1)}}.
\]
Thus the width is bounded independently of the input size, but doubly exponentially in $1/\eps$: the exponent is exponential in $1/\eps$, whereas the exponent of $B_*$ in~\eqref{eq:widthcap} is polynomial in $1/\eps$. The loss comes from the estimate $D\ge\eps\OPT$, which ignores that the cut optimum edges also increase $D$.

\subsection{Using the growth of the allowance}\label{sec:potential}

Section~\ref{sec:firstbound} bounded the growth of $L(B)$ in every paid step by the newly cut cost relative to $D$, and then replaced $D$ by its initial value $\eps\OPT$. But the allowance grows, and it grows by the same cuts: by~\eqref{eq:Dupdate}, every newly cut edge use of $\cPs$ increases $D$ by $\alpha-1$ times its cost. In a paid step, Lemma~\ref{lem:logstep} and~\eqref{eq:Dupdate} give
\[
 \frac{L(B')}{L(B)}\le1+\frac{8(\alpha+1)c(E_P)}{D}+\frac{8(\alpha+1)c(E_Q)}{AD},
 \qquad
 \frac{D'}{D}=1+\frac{(\alpha-1)c(E_P)}{D}.
\]
Apart from the term caused by $\cQ$, both factors are determined by the same ratio $c(E_P)/D$. As $D$ grows, the same cut cost gives a smaller ratio, so the payments made in later steps allow less growth of $L(B)$.

The two factors differ only in their constants $8(\alpha+1)$ and $\alpha-1$, and a power of $D'/D$ makes up for this difference. Its exponent is the ratio of the two constants, rounded up,
\[
 p=\left\lceil\frac{8(\alpha+1)}{\alpha-1}\right\rceil,
\]
as defined in~\eqref{eq:widthcap}. Since $p(\alpha-1)\ge8(\alpha+1)$, Bernoulli's inequality gives
\begin{equation}\label{eq:bernoulli}
 \left(\frac{D'}D\right)^p
 =\left(1+\frac{(\alpha-1)c(E_P)}{D}\right)^p
 \ge1+\frac{p(\alpha-1)c(E_P)}{D}
 \ge1+\frac{8(\alpha+1)c(E_P)}{D}.
\end{equation}
So, apart from the term caused by $\cQ$, $D^p$ grows at least by the factor by which $L(B)$ grows, and the quotient $L(B)/D^p$ does not increase. Over the whole process, $D$ grows from $\eps\OPT$ to at most $(\alpha-1+\eps)\OPT$, by~\eqref{eq:allowance}, since the multiset $K$ of edge uses of $\cPs$ between different parts costs at most $\OPT$. So, up to a constant factor caused by $\cQ$, $L(B)$ grows at most by the factor $(1+(\alpha-1)/\eps)^p$, which is polynomial in $1/\eps$, and $B$ is singly exponential in a polynomial in $1/\eps$, rather than doubly exponential as in Section~\ref{sec:firstbound}; we prove this below. To this end, define the \emph{potential}
\[
 \Psi(B,D)=L(B)\left(\frac{\eps\OPT}{D}\right)^p,
\]
normalized so that its initial value is $L(B_0)$, since initially $D=\eps\OPT$. The next lemma shows that no step increases it, except for a small factor caused by the cut edges of $\cQ$.

\begin{lemma}[potential growth]\label{lem:potential}
In every step,
\begin{equation}\label{eq:potentialstep}
 \Psi(B',D')\le\Psi(B,D)
 \exp\left(\frac{8(\alpha+1)c(E_Q)}{A\eps\OPT}\right).
\end{equation}
Consequently, $B<B_*$ throughout the process.
\end{lemma}

\begin{proof}
In a free step, $B'=B$ and $D'\ge D$, so the potential does not increase. In a paid step, the bound on $L(B')/L(B)$ above and $1+x+y\le(1+x)(1+y)$ for $x,y\ge0$ give
\[
 \frac{L(B')}{L(B)}
 \le\left(1+\frac{8(\alpha+1)c(E_P)}{D}\right)\left(1+\frac{8(\alpha+1)c(E_Q)}{AD}\right).
\]
By~\eqref{eq:bernoulli}, the first factor is at most $(D'/D)^p$. By $1+x\le\ee^x$ and $D\ge\eps\OPT$, the second factor is at most $\exp(8(\alpha+1)c(E_Q)/(A\eps\OPT))$. Hence
\[
 \frac{\Psi(B',D')}{\Psi(B,D)}
 =\frac{L(B')}{L(B)}\left(\frac{D}{D'}\right)^p
 \le\exp\left(\frac{8(\alpha+1)c(E_Q)}{A\eps\OPT}\right),
\]
which is~\eqref{eq:potentialstep}.

Multiplying~\eqref{eq:potentialstep} over all steps and using Lemma~\ref{lem:fresh} and~\eqref{eq:A}, we obtain
\[
 \Psi(B,D)
 \le L(B_0)\exp\left(\frac{8(\alpha+1)}{A\eps\OPT}\sum c(E_Q)\right)
 \le L(B_0)\exp\left(\frac{8(\alpha+1)\beta}{A\eps}\right)
 \le\ee L(B_0),
\]
since $8(\alpha+1)\beta/(A\eps)\le\beta/(1+\beta)<1$. Finally, $D\le(\alpha-1+\eps)\OPT$, as noted before the lemma. Using $\ln B<L(B)$, $L(B_0)=\ln(2AB_0)$, $\ee<3$, and the definition~\eqref{eq:widthcap} of $B_*$, we obtain
\[
 \ln B<L(B)=\Psi(B,D)\left(\frac{D}{\eps\OPT}\right)^p
 \le\ee\ln(2AB_0)\left(1+\frac{\alpha-1}{\eps}\right)^p
 <3\left(1+\frac{\alpha-1}{\eps}\right)^p\ln(2AB_0)\le\ln B_*.
\]
\end{proof}

\subsection{Proof of the structural theorem}\label{sec:structure-proof}

\begin{proof}[Proof of Theorem~\ref{thm:process}]
By Lemma~\ref{lem:refinement-step}, every step divides at least one part and preserves the invariant~\eqref{eq:treewidth}. Single vertices have diameter zero, so the stopping condition holds at the latest when all parts are single vertices; hence the process stops after at most $n-1$ steps, and the leaves of the final tree satisfy the stopping condition~\eqref{eq:goal}. By the invariant~\eqref{eq:treewidth} and Lemma~\ref{lem:potential}, every node $W$ of this tree satisfies $w(W)\le B<B_*$.
\end{proof}

\section{Unweighted digraphs and the symmetric case}\label{sec:applications}

This section adapts the algorithm and its analysis to unweighted digraphs, which proves Theorem~\ref{thm:graph-path}, and then derives the symmetric case. Table~\ref{tab:inputs} lists the constants of their reference families, which Appendix~\ref{sec:initialization} constructs.

\subsection{Unweighted digraphs}\label{sec:graphic}

Here the instance is given by a strongly connected unweighted digraph $G=(V,E)$, and $c$ is its graph metric: $c(u,v)$ is the length of a shortest $u$-$v$ path in $G$. The given algorithm is now an $\alpha$-approximation algorithm for Asymmetric Graph TSP, so it may only be called on graph metrics of unweighted digraphs. The local candidate of Section~\ref{sec:candidate} calls it on the restriction of $c$ to a vertex set $U$. This restriction need not be the graph metric of an unweighted digraph, because a shortest path in $G$ between two vertices of $U$ may leave $U$. We therefore measure distances inside $U$ in the induced subgraph $G[U]$, and call the algorithm for Asymmetric Graph TSP on $G[U]$. This is possible only if $G[U]$ is strongly connected, and we then call $U$ \emph{strongly connected}; a single vertex is strongly connected. So local candidates, and with them the distances inside a set, are needed only for strongly connected sets. Sets that are not strongly connected are divided, both in the algorithm and in the analysis: the dynamic program handles them only by combining smaller sets, and the refinement process divides them first, as described below. For a strongly connected set $U$ and $u,v\in U$, let the \emph{internal distance} $\operatorname{dist}_U(u,v)$ be the length of a shortest $u$-$v$ path in $G[U]$, and let $d_U=\max_{u,v\in U}\operatorname{dist}_U(u,v)$ be the \emph{internal diameter} of $U$; it replaces the diameter of Section~\ref{sec:candidate} throughout. We go through the changes in the order of Sections~\ref{sec:prelim} to~\ref{sec:width}; everything not mentioned carries over unchanged.

\paragraph{Walks in $G$ (Section~\ref{sec:prelim}).}
Since $c$ is the graph metric of $G$, we work with walks in $G$: $\cPs$ is an optimum feasible family of walks in $G$, and the reference family $\cQ$ also consists of walks in $G$ (Appendix~\ref{sec:initialization}). Every edge use of $\cPs$ or $\cQ$ between two vertices of a set $U$ is then an edge of $G[U]$. The walks of $\cQ$ need not be vertex-disjoint paths, so the order $\sigma$ of Section~\ref{sec:interface} is obtained by traversing them one after another, in a fixed order, and listing each vertex when it is visited for the first time. Lemma~\ref{lem:first-order} then holds with at most $r_{\cQ}(U)$ intervals. Indeed, while the traversal runs through a fragment of $\cQ$ in $U$, it lists exactly the vertices of $U$ that are first visited in this fragment, one after another, so they form an interval of $\sigma$. Every vertex of $U$ is first visited in some fragment of $\cQ$ in $U$, so $U$ is the union of at most $r_{\cQ}(U)$ such intervals.

\paragraph{Local candidates (Section~\ref{sec:local}).}
For a strongly connected set $U$, the local candidate calls the algorithm for Asymmetric Graph TSP on $G[U]$ and replaces each edge of the returned tour by a shortest path in $G[U]$. This gives a closed walk $C_U$ in $G[U]$ that visits all vertices of $U$ and costs at most $\alpha$ times the minimum cost of such a walk. The local candidate connects the pairs by shortest paths in $G[U]$ instead of edges; each of these paths costs at most $d_U$, so the cost bound~\eqref{eq:candidatecost} of the local candidate still holds. The proof of Lemma~\ref{lem:localbound} applies with internal distances: the fragments of $\cPs$ in $U$ are walks in $G[U]$, and linking them by shortest paths in $G[U]$ gives a closed walk in $G[U]$ that visits all vertices of $U$ and costs at most $c(\cPs[U])+r_{\cPs}(U)d_U$. Hence Lemma~\ref{lem:hybrid} holds for partitions into strongly connected sets. A set that is not strongly connected has no local candidate.

\paragraph{The dynamic program (Section~\ref{sec:dp}).}
Algorithm~\ref{alg:main} records local candidates only for strongly connected sets; the families kept for the other sets arise only by combining smaller sets. The rest of the algorithm, including the guesses of edges between two sets, is unchanged, and the cell $Z[V,M_0]$ is not empty, because $V$ is strongly connected. Theorem~\ref{thm:dp} holds for refinement trees with strongly connected leaves, since its proof uses local candidates only at the leaves.

\paragraph{Dividing along levels (Section~\ref{sec:fragments}).}
The heights use internal distances: for a strongly connected set $U$, choose $(a_U,b_U)$ with $\operatorname{dist}_U(a_U,b_U)=d_U$ and set $f_U(v)=\operatorname{dist}_U(a_U,v)$. For an edge $(u,v)$ of $G[U]$, a shortest path in $G[U]$ from $a_U$ to $u$, followed by $(u,v)$, shows that $f_U(v)-f_U(u)\le1=c(u,v)$. The edge uses of $\cPs$ and $\cQ$ inside $U$ are edges of $G[U]$, so the proof of Lemma~\ref{lem:coarea} applies to them. The rest of Section~\ref{sec:fragments} does not depend on how the heights are defined.

\paragraph{The process and its payments (Sections~\ref{sec:process} and~\ref{sec:width}).}
Heights and local candidates need strongly connected sets, so the process first divides parts that are not. At the beginning of each step, if some part $U$ is not strongly connected, choose $v\in U$ such that the set $U_1$ of vertices reachable from $v$ in $G[U]$ is not all of $U$, which is possible because $G[U]$ is not strongly connected, and divide $U$ into $U_1$ and $U_2=U\setminus U_1$. This is a free split in the sense of Section~\ref{sec:overview}, with $U_1$ in the role of the lower set: no edge of $G[U]$ leaves $U_1$, so every fragment of $\cPs$ or $\cQ$ in $U$ consists of a part in $U_2$ followed by a part in $U_1$, either of which may be empty. Hence neither set has more fragments of $\cPs$ or of $\cQ$ than $U$, and the division preserves the invariant~\eqref{eq:treewidth}. Such a step is a free step. The stopping condition is tested only when all parts are strongly connected, so the final parts are strongly connected. With these changes, Lemma~\ref{lem:refinement-step} holds with the same proof.

Internal distances are taken inside the current part, so a child may have a larger internal diameter than its parent. The analysis does not need monotonicity: the stopping condition and the payment of each step involve only the current parts and their height functions, and the process terminates because every step divides a part. Section~\ref{sec:width} carries over verbatim.

The \emph{version of Theorem~\ref{thm:unified} for unweighted digraphs} states that an $\alpha$-approximation algorithm for Asymmetric Graph TSP yields an $(\alpha+\eps)$-approximation algorithm for Asymmetric Multi-Path TSP with $k$ prescribed pairs on unweighted digraphs. For $k=1$, this is Theorem~\ref{thm:graph-path}.

\begin{proof}[Proof of Theorem~\ref{thm:unified} for unweighted digraphs]
Lemma~\ref{lem:single-reference} computes, in polynomial time, a reference family of walks in $G$ with $\beta=\alpha(2k+1)$ and $\ell=\lceil\alpha(k+1)\rceil$. Run Algorithm~\ref{alg:main} with this family, the changes above, and the bound $B_*$ of~\eqref{eq:widthcap}. By the preceding discussion, the arguments of Sections~\ref{sec:local} to~\ref{sec:width} give a refinement tree whose nodes have width less than $B_*$ and whose leaves are strongly connected and satisfy the stopping condition~\eqref{eq:goal}. By Theorem~\ref{thm:dp} and Lemma~\ref{lem:hybrid}, the algorithm returns a feasible family of cost at most $(\alpha+\eps)\OPT$. The dynamic program calls the algorithm for Asymmetric Graph TSP only on $G[U]$ for strongly connected sets $U$. The number of oracle calls and the additional work satisfy the same degree bound as in the proof of Theorem~\ref{thm:unified}.
\end{proof}

Using the algorithm of Vygen~\cite{Vyg26} with error $\eps/2$, a $(3+2\sqrt2+\eps/2)$-approximation algorithm for Asymmetric Graph TSP, Theorem~\ref{thm:graph-path} with error $\eps/2$ gives the ratio $3+2\sqrt2+\eps$ stated after it.

The assumption that $G$ is strongly connected is no restriction for Theorem~\ref{thm:graph-path}. An $s$-$t$ walk that leaves a strongly connected component of $G$ cannot return to it, so a feasible walk visits the components in a fixed order, from the component of $s$ to that of $t$, with one entry and one exit vertex in each component, where $s$ is the entry of the first and $t$ the exit of the last. For each component and each choice of entry and exit vertex, compute an approximate solution inside the component, by the algorithm of Theorem~\ref{thm:graph-path} if the two vertices differ and by the given algorithm if they agree; then choose the entry and exit vertices along this order, joined by arcs of $G$, by a shortest-path computation. The arcs between components are paid for exactly, so the guarantee $\alpha+\eps$ carries over.

\subsection{Symmetric instances}

\begin{corollary}[symmetric prescribed pairs]\label{cor:symmetric}
Let $\alpha>1$, $\eps>0$, and $k\ge1$ be constants. An $\alpha$-approximation algorithm for metric TSP yields an $(\alpha+\eps)$-approximation algorithm for Multi-Path TSP with $k$ prescribed pairs. The statement also holds with both problems restricted to unweighted graphs.
\end{corollary}

\begin{proof}
A symmetric metric is a directed metric, and the local candidate calls the given algorithm for metric TSP and orients the returned tour along its traversal. Lemma~\ref{lem:undirected-seed} supplies $\cQ$ with $\beta=2$ and $\ell=k$, and Lemma~\ref{lem:general} gives the result. For an unweighted graph, apply Section~\ref{sec:graphic} to the digraph that has both orientations of every edge. A set is then strongly connected exactly when it induces a connected subgraph, so the given algorithm, here one for metric TSP on unweighted graphs, is always called on a connected unweighted graph, and Lemma~\ref{lem:undirected-seed} supplies $\cQ$ as walks in the given graph.
\end{proof}

For one pair this gives another reduction with the guarantee of Traub, Vygen and Zenklusen~\cite{TVZ22}, and for a fixed number of pairs one with the guarantee of~\cite[Corollary~16.25]{TV25}.

\section{Conclusion}

We have shown that, for all constants $\alpha>1$ and $\eps>0$, any $\alpha$-approximation algorithm for Asymmetric TSP yields an $(\alpha+\eps)$-approximation algorithm for Asymmetric Path TSP. The same dynamic program handles a fixed number of prescribed pairs, unweighted digraphs, and the symmetric case; for Asymmetric Graph Path TSP, it gives a $(3+2\sqrt2+\eps)$-approximation algorithm.

The main conceptual point is a new use of the guessed edges in dynamic programs of this kind: besides improving the cost bound, their cost bounds the number of edges that the dynamic program has to guess. This replaces iterative improvement by refinement, and it may be useful for other problems in which such dynamic programs are used.

The analogous question for integrality ratios, whether the natural LP relaxations of Asymmetric TSP and Asymmetric Path TSP have the same integrality ratio, remains open; this is Open Problem~9.25 of~\cite{TV25}. The best known relation, due to K\"ohne, Traub and Vygen~\cite{KTV20}, is that if the integrality ratio for Asymmetric TSP is $\rho$, then the one for the path version is at most $4\rho-3$. For unweighted undirected graphs, the analogous question has a negative answer: the integrality ratio is at most $\frac75$ for the tour version and exactly $\frac32$ for the path version~\cite{SV14}, whereas the approximation ratios agree up to~$\eps$ by~\cite{TVZ22} and Corollary~\ref{cor:symmetric}.

\section*{Statement on the use of AI}
Large language models (ChatGPT~6 Astra by OpenAI and Claude Opus~5.5 by Anthropic) were used to assist in exploring proof ideas and in writing this paper. The main direction originated with the author: using finer partitions to control the local errors, dividing vertex sets by the heights of their vertices, and relating the integral of the upward crossing numbers over all levels to the cost of the walks, with the aim of bounding the numbers of fragments and hence the size of the dynamic program. Building on this direction, the specific refinement process and its analysis by a potential were developed in discussions with the models. In addition, ChatGPT~6.2 Astra by OpenAI was used to formalize Theorems~\ref{thm:main} to~\ref{thm:graph-path} in Lean~4; the formal proofs are checked by the Lean kernel (see footnote~\ref{fn:lean}). All arguments in this paper were verified by the author, who takes full responsibility for its mathematical content.

\begingroup\small
\setlength{\bibsep}{1.5pt plus 1pt minus 1pt}

\endgroup

\clearpage
\appendix

\section{Constructing the reference family}\label{sec:initialization}

The algorithm needs a reference family $\cQ$ that satisfies~\eqref{eq:qref}: its walks together visit every vertex, there are at most $\ell$ of them, and they cost at most $\beta\OPT$. The family need not be a feasible solution, since its walks may start and end anywhere. For Asymmetric Path TSP, the Feige--Singh reduction provides $\cQ$ (Section~\ref{sec:interface}); this appendix computes $\cQ$ in polynomial time for the other problems of Table~\ref{tab:inputs}. If all distances are zero, any path through all vertices is a reference family; so we assume throughout that some distance is positive.

\subsection{Directed metrics}\label{sec:seed-metric}

For a directed metric, we construct a family of vertex-disjoint paths that satisfies~\eqref{eq:qref} with
\begin{equation}\label{eq:seed-bounds}
 \beta=\alpha(2k+1),\qquad
 \ell=\lceil\alpha(k+1)\rceil.
\end{equation}

For $\tau>0$, the \emph{capped distances} $c_\tau(u,v)=\min\{c(u,v),\tau\}$ again form a directed metric: if $c_\tau(u,v)$ or $c_\tau(v,w)$ equals $\tau$, then $c_\tau(u,w)\le\tau\le c_\tau(u,v)+c_\tau(v,w)$, and otherwise the triangle inequality for $c$ applies. Call the algorithm for Asymmetric TSP on $c_\tau$, which returns a tour $T_\tau$ of $V$, and delete the edges $e$ of $T_\tau$ with $c(e)\ge\tau$, or one edge if there is no such edge. The rest of $T_\tau$ is a family $\cQ_\tau$ of vertex-disjoint paths that together visit every vertex, one path for each deleted edge, and every edge $e$ of $\cQ_\tau$ has $c(e)=c_\tau(e)$.

The choice of $\tau$ balances cost against the number of paths. Join $t_i$ to $s_{i+1}$, where $s_{k+1}=s_1$, by the edge $(t_i,s_{i+1})$ if the two vertices differ; together with the walks of $\cPs$, these edges form a closed walk through all vertices of capped cost at most $\OPT+k\tau$. Shortcutting it gives a tour of $V$, so
\begin{equation}\label{eq:capped-cost}
 c(\cQ_\tau)\le c_\tau(T_\tau)\le\alpha(\OPT+k\tau).
\end{equation}
Each deleted edge $e$ with $c(e)\ge\tau$ has capped cost $\tau$, so $\cQ_\tau$ has at most $\max\{1,\alpha(\OPT/\tau+k)\}$ paths. A small $\tau$ keeps the cost small, and a large $\tau$ keeps the number of paths small; for $\OPT\le\tau<2\OPT$, both bounds match~\eqref{eq:seed-bounds}. The algorithm does not know $\OPT$, so it tries every $\tau=2^jc(u,v)$ with $c(u,v)>0$ and $0\le j\le\lceil\log_2(n+k)\rceil$, discards the families with more than $\ell$ paths, and returns a cheapest remaining family $\cQ$.

\begin{lemma}[capped distances]\label{lem:capped}
For every directed metric, the returned family $\cQ$ satisfies~\eqref{eq:qref} with the bounds~\eqref{eq:seed-bounds}. The construction makes at most $n^2(\lceil\log_2(n+k)\rceil+1)$ calls to the algorithm for Asymmetric TSP, each on $n$ vertices, and performs polynomial additional work.
\end{lemma}

\begin{proof}
Suppose first that $\OPT>0$. Some optimum solution has at most $n+k$ edges, since shortcutting a vertex that lies on several of its paths, except where it is an end vertex, does not increase the cost. Let $e$ be an edge of maximum cost in such a solution; then $0<c(e)\le\OPT\le(n+k)c(e)$. So the smallest $j\ge0$ with $2^jc(e)\ge\OPT$ satisfies $j\le\lceil\log_2(n+k)\rceil$, and the tried value $\tau=2^jc(e)$ satisfies $\OPT\le\tau<2\OPT$. For this $\tau$, $\cQ_\tau$ has at most $\max\{1,\alpha(k+1)\}\le\ell$ paths and, by~\eqref{eq:capped-cost}, costs at most $\alpha(\OPT+k\tau)\le\alpha(2k+1)\OPT$. Suppose next that $\OPT=0$, and let $\tau$ be the smallest positive distance, which is tried. Then every edge of $T_\tau$ of positive cost has $c(e)\ge\tau$ and is deleted, so $\cQ_\tau$ costs zero, and by the bound on the number of paths with $\OPT=0$, it has at most $\max\{1,\alpha k\}\le\ell$ paths. In both cases, this family is not discarded, and the returned family costs no more. There are at most $n(n-1)$ positive distances and $\lceil\log_2(n+k)\rceil+1$ values of $j$.
\end{proof}

\subsection{Unweighted digraphs}\label{sec:seed-one-guess}

For an unweighted digraph $G$, we follow Appendix~\ref{sec:seed-metric}, but capped distances need not be the graph metric of an unweighted digraph, so we add new paths instead. For an integer $\tau\ge1$ and each $i$ with $t_i\ne s_{i+1}$, add an \emph{artificial path}: a directed path of $\tau$ edges from $t_i$ to $s_{i+1}$, in this direction only, whose internal vertices are new and used by no other artificial path. Let $G_\tau$ be the resulting unweighted digraph. For $\tau=1$, an artificial path is the edge $(t_i,s_{i+1})$, which is not added if it is already present. The walks of $\cPs$, joined by the artificial paths, form a closed walk of cost at most $\OPT+k\tau$ that visits every vertex of $G_\tau$; in particular, $G_\tau$ is strongly connected. Call the algorithm for Asymmetric Graph TSP on $G_\tau$, obtaining, as in Section~\ref{sec:graphic}, a closed walk $C$ in $G_\tau$ that visits all its vertices. Each internal vertex of an artificial path has one entering and one leaving edge, so whenever $C$ enters an artificial path, it traverses the whole path from $t_i$ to $s_{i+1}$, at cost $\tau$. Deleting these traversals takes the place of deleting the edges of capped cost $\tau$: reading $C$ cyclically, the pieces between consecutive traversals form a family $\cQ_\tau$ of walks in $G$, one for each traversal and some possibly of length zero, that together visit every vertex of $V$; if $C$ traverses no artificial path, let $\cQ_\tau=\{C\}$. As in Section~\ref{sec:graphic}, these walks need not be vertex-disjoint paths.

As in Appendix~\ref{sec:seed-metric}, we need a value $\tau$ with $\OPT\le\tau<2\OPT$. If every prescribed pair has equal end vertices and these vertices cover $V$, then $\OPT=0$, and the zero-length walks at these vertices form a reference family. Otherwise $\OPT\ge1$, since distances are integers. Moreover, as in the proof of Lemma~\ref{lem:capped}, some optimum solution has at most $n+k$ edges of the metric $c$, and each of them costs at most $n-1$, since $G$ is strongly connected; so $\OPT\le N$ for $N=(n+k)(n-1)$. Hence it suffices to compute $\cQ_\tau$ for $\tau=2^j$ with $0\le j\le\lceil\log_2N\rceil$; discard the families with more than $\ell$ walks, and return a cheapest remaining family $\cQ$.

\begin{lemma}[a reference family for unweighted digraphs]\label{lem:single-reference}
For every strongly connected unweighted digraph $G$, the returned family $\cQ$ consists of walks in $G$ and satisfies~\eqref{eq:qref} with the bounds~\eqref{eq:seed-bounds}. The construction makes $O(\log(n+k))$ calls to the algorithm for Asymmetric Graph TSP, on unweighted digraphs with $O(n+k(n+k)n)$ vertices, and performs polynomial additional work.
\end{lemma}

\begin{proof}
If $\OPT=0$, the returned family consists of $k\le\ell$ walks of length zero. Otherwise, the proof of Lemma~\ref{lem:capped} applies, with the following changes. By the closed walk above, $c(\cQ_\tau)\le c(C)\le\alpha(\OPT+k\tau)$, which replaces~\eqref{eq:capped-cost}. The walk $C$ traverses artificial paths at most $c(C)/\tau\le\alpha(\OPT/\tau+k)$ times, so $\cQ_\tau$ has at most $\max\{1,\alpha(\OPT/\tau+k)\}$ walks; since walks of length zero are kept, every vertex of $V$ is still visited. Since $1\le\OPT\le N$, some tried value satisfies $\OPT\le\tau<2\OPT$. The largest value of $\tau$ is at most $2N$, so the artificial paths have at most $2kN$ internal vertices in total.
\end{proof}

\subsection{Undirected instances}\label{sec:seed-undirected}

For undirected instances, a minimum spanning tree gives a reference family with $\beta=2$ and $\ell=k$.

\begin{lemma}[an undirected reference family]\label{lem:undirected-seed}
For an instance of Multi-Path TSP in a connected undirected graph with nonnegative edge costs, one can compute in polynomial time a family of at most $k$ closed walks that visits every vertex and has total cost at most $2\OPT$.
\end{lemma}

\begin{proof}
Add a new vertex $z$, and join it by edges of cost zero to the distinct vertices among $s_1,\ldots,s_k$. The walks of an optimum solution, together with these edges, form a connected subgraph of the new graph that contains all its vertices, so a minimum spanning tree $T$ of the new graph costs at most $\OPT$. Deleting $z$ from $T$ leaves at most $k$ trees that together contain $V$. Double the edges of each tree and traverse it as a closed walk; a single vertex gives a zero-length walk. These at most $k$ closed walks visit every vertex and cost at most $2\OPT$.
\end{proof}

In a symmetric metric, shortcutting these closed walks, keeping the first occurrence of every vertex, turns them into vertex-disjoint paths without increasing the cost. For an unweighted graph, the walks are kept, oriented along their traversals.


\begin{thebibliography}{99}

\bibitem{AGMOS17}
Arash Asadpour, Michel X. Goemans, Aleksander M\k{a}dry, Shayan Oveis Gharan, and Amin Saberi.
An $O(\log n/\log\log n)$-approximation algorithm for the asymmetric traveling salesman problem.
\emph{Operations Research}, 65(4):1043--1061, 2017.

\bibitem{BCKLMM07}
Avrim Blum, Shuchi Chawla, David R. Karger, Terran Lane, Adam Meyerson, and Maria Minkoff.
Approximation algorithms for orienteering and discounted-reward {TSP}.
\emph{SIAM Journal on Computing}, 37(2):653--670, 2007.

\bibitem{CP07}
Chandra Chekuri and Martin P\'al.
An $O(\log n)$ approximation ratio for the asymmetric traveling salesman path problem.
\emph{Theory of Computing}, 3:197--209, 2007.

\bibitem{CT26}
Manuel Christalla and Vera Traub.
Approximation algorithms for asymmetric multi-path TSP.
arXiv:2611.26188, 2026.

\bibitem{FS07}
Uriel Feige and Mohit Singh.
Improved approximation ratios for traveling salesperson tours and paths in directed graphs.
In \emph{Approximation, Randomization, and Combinatorial Optimization (APPROX/RANDOM)}, volume 4627 of \emph{Lecture Notes in Computer Science}, pages 104--118. Springer, 2007.

\bibitem{Fri13}
Zachary Friggstad.
Multiple traveling salesmen in asymmetric metrics.
In \emph{Approximation, Randomization, and Combinatorial Optimization (APPROX/RANDOM)}, volume 8096 of \emph{Lecture Notes in Computer Science}, pages 173--188. Springer, 2013.

\bibitem{HY26}
MohammadTaghi Hajiaghayi and Changki Yun.
A common improved bound for asymmetric TSP and path TSP.
arXiv:2611.18498, 2026.

\bibitem{KKO24}
Anna R. Karlin, Nathan Klein, and Shayan Oveis Gharan.
A (slightly) improved approximation algorithm for metric TSP.
\emph{Operations Research}, 72(6):2543--2594, 2024.

\bibitem{KTV20}
Anna K\"ohne, Vera Traub, and Jens Vygen.
The asymmetric traveling salesman path LP has constant integrality ratio.
\emph{Mathematical Programming}, 183:379--395, 2020.

\bibitem{LN08}
Fumei Lam and Alantha Newman.
Traveling salesman path problems.
\emph{Mathematical Programming}, 113:39--59, 2008.

\bibitem{Li26}
Hong Li.
Reducing prize-collecting stroll and related routing problems to prize-collecting {TSP}.
arXiv:2606.18157, 2026.

\bibitem{SV14}
Andr\'as Seb\H{o} and Jens Vygen.
Shorter tours by nicer ears: 7/5-approximation for the graph-{TSP}, 3/2 for the path version, and 4/3 for two-edge-connected subgraphs.
\emph{Combinatorica}, 34(5):597--629, 2014.

\bibitem{Sve15}
Ola Svensson.
Approximating ATSP by relaxing connectivity.
In \emph{Proceedings of the 56th Annual IEEE Symposium on Foundations of Computer Science (FOCS)}, pages 1--19, 2015.

\bibitem{STV20}
Ola Svensson, Jakub Tarnawski, and L\'aszl\'o A. V\'egh.
A constant-factor approximation algorithm for the asymmetric traveling salesman problem.
\emph{Journal of the ACM}, 67(6):37:1--37:53, 2020.

\bibitem{Tra17}
Vera Traub.
Approximating the $s$-$t$-path {TSP}.
Master's thesis, University of Bonn, 2017.

\bibitem{TV19a}
Vera Traub and Jens Vygen.
Approaching $\frac32$ for the $s$-$t$-path TSP.
\emph{Journal of the ACM}, 66(2):14:1--14:17, 2019.

\bibitem{TV22}
Vera Traub and Jens Vygen.
An improved approximation algorithm for the asymmetric traveling salesman problem.
\emph{SIAM Journal on Computing}, 51(1):139--173, 2022.

\bibitem{TV25}
Vera Traub and Jens Vygen.
\emph{Approximation Algorithms for Traveling Salesman Problems}.
Cambridge University Press, 2025.

\bibitem{TVZ22}
Vera Traub, Jens Vygen, and Rico Zenklusen.
Reducing path TSP to TSP.
\emph{SIAM Journal on Computing}, 51(3):STOC20-24--STOC20-53, 2022.

\bibitem{Vyg26}
Jens Vygen.
Better approximation guarantee for asymmetric TSP.
arXiv:2603.14334, 2026.

\bibitem{Zen19}
Rico Zenklusen.
A 1.5-approximation for path TSP.
In \emph{Proceedings of the 30th Annual ACM-SIAM Symposium on Discrete Algorithms (SODA)}, pages 1539--1549, 2019.

\end{thebibliography}
\end{document}